\documentclass[twocolumn]{article} 
\usepackage{graphicx} 
\usepackage[margin=20truemm]{geometry}
\usepackage[version=4]{mhchem}
\usepackage{mathtools}
\usepackage{nicefrac}
\usepackage{amsmath,amsfonts,amssymb,epsfig,graphicx,fancyhdr,subcaption,epstopdf,hyperref,siunitx,xcolor,wasysym, comment}
\usepackage{bm}
\usepackage{color}
\usepackage{here}
\usepackage{autobreak}
\usepackage{authblk}

\usepackage{caption}
\title{
Time delay as the origin of oscillations in anodic Si electrodissolution\\
}
\author{Yukiteru Murakami$^*$}
\author{Katharina Krischer$^{\dagger}$} 

\affil{School of Natural Sciences, Physics Department, Nonequilibrium Chemical Physics,\\ Technische Universität München, James-Franck-Str. 1, D-85748 Garching, Germany\\\textit{*ga92dom@mytum.de, $^{\dagger}$krischer@tum.de}}
\date{}

\begin{document}
\twocolumn[
 \maketitle
    \begin{abstract}
        Silicon is the most important semiconductor electrode with applications in photoelectrochemistry and sensor technology. Yet its electrochemistry exhibits many poorly understood phenomena, including oscillations during the anodic dissolution of silicon electrodes. 
        In this article, we present a mathematical model based on physicochemical steps that captures these oscillations and enables a thorough understanding of the underlying mechanism. The model describes the formation and dissolution of an oxide layer, and determines the oxide composition and the electrostatic potential in the direction perpendicular to the electrode.
        Oscillations occur if the following conditions are fulfilled:
        1. The etching speed increases with defects in the oxide layer
        2. The number of defects decreases with increasing electric field strength at the Si-oxide interface.
        3. There is a sufficient time delay between the production and the etching of the oxide.
        Numerical simulations reproduce experimental results well, including the dependence of the oscillation amplitude and period on the potential, as well as the hysteresis behavior observed in cyclic voltammetry. Based on these results, we derive a simplified time-delay differential equation model. Using linear stability analysis, we confirm the essential role of the time delay for the oscillatory oxide-layer dynamics. The basic steps of the model are general and in line with the point defect model for growth and dissolution of passive films on metal electrodes. Therefore, it is likely applicable to a variety of oscillating anodic metal or semiconductor dissolution reactions. \\    
    \end{abstract}
    \vspace{1 em}      
]

\section{Introduction}
Because of its importance in the electronics industry and its application in electrochemical devices, such as sensors or photoelectrochemical cells, silicon is arguably the semiconductor material whose electrochemistry has been most thoroughly studied among all semiconductors \cite{Zhang2001}. Yet, there are many fundamental processes occurring at the electrified Si/electrolyte interface that are still only poorly understood. Here the situation seems little changed from twenty years ago, when Macdonald, in the preface to the book "Electrochemistry of Si and its oxides", \cite{Zhang2001} pointed out that: " ...\textit{ the lack of a comprehensive account of the electrochemistry of silicon in aqueous solution at the fundamental level is surprising and troubling […] some of the electrochemical properties of this element are not as well known as might be warranted by its importance in a modern technological society.}"

One of these still puzzling, yet intensively studied phenomena are self-sustained oscillations in current or potential in a fluoride-containing electrolyte, first reported nearly 70 years ago by Turner \cite{Turner1958}.
The last decades revealed that these oscillations are connected to a variety of further nonlinear phenomena, such as birhythmicity \cite{Wiehl2021} and even bichaoticity \cite{Tosolini2019}, i.e. the coexistence of different types of stable oscillations or chaotic attractors, at the same parameter values, or the formation of spatio-temporal patterns \cite{Miethe2009, Schoenleber2014}. 
Two notable examples of the latter are chimera states, i.e., synchronization patterns in ensembles of identical oscillators or uniform oscillating media that are composed of a synchronously and an incoherently oscillating part \cite{Schmidt2014}, and frequency clusters \cite{Patzauer2021}, that are composed of a few regions that oscillate with distinct frequencies. 
In both cases, Si electrooxidation is one of the few experimental systems where the spontaneous formation of such patterns has been observed, making the ‘Si oscillator’ a prototypical system for the experimental exploration of pattern formation or the validation of theories. 
Understanding the mechanism of the current oscillations during electrodissolution of Si is therefore important both for our knowledge of the dynamics of nonlinear systems and industrial applications where Si electrochemistry is employed. 

As mentioned above, the study of Si oscillations has a long history. 
Already, Turner pointed out that during the oscillations, a passivating silicon oxide film covers the electrode surface. The oxide film is formed electrochemically and chemically etched by fluoride species according to reactions (\ref{eq:ox}) and (\ref{eq:etch_overall}), respectively:
\begin{align}
\rm Si + 2H_2O + \lambda_{VB} h^+  \xrightarrow \rm SiO_2 + 4H^+ + (4-\lambda_{VB}) e^{-}, \tag{R0a} \label{eq:ox}
\end{align}
\begin{align}
\rm SiO_2 + 6HF \xrightarrow \rm SiF_6^{2-} + 2H_2O + 2H^{+}, \tag{R0b} \label{eq:etch_overall}
\end{align}
where $1 \leq \lambda_{\rm VB} \leq 4$ is the number of the valency band holes taking part in the oxidation reaction \cite{Cattarin2000}. 
Since holes are required for the oxidation process, the reaction (\ref{eq:ox}) proceeds with either p-type silicon or illuminated n-type silicon.
Under potentiostatic conditions, the oscillatory oxidation current is observed over a wide range of parameters.

From the outset, it was speculated that the oscillations are caused by some kind of interplay between oxide film formation and dissolution. This triggered a large number of investigations into the film properties during oscillations using various in-situ and ex-situ techniques, such as IR spectroscopy \cite{Chazalviel1998,Cattarin1998a}, ellipsometry \cite{Aggour1995} atomic force microscopy \cite{Lehmann1996}, x-ray reflectometry \cite{Lehmann1996}, microwave reflectivity \cite{Cattarin1998a} and transmission electron microscopy \cite{Aggour1995}. An overview of experimental studies on current oscillations up to 2001 can be found in chapter 5 of \cite{Zhang2001}. The studies showed that many properties of the oxide film oscillated along with the current or potential, most notably the thickness of the oxide layer \cite{Chazalviel1998}. Furthermore, it was observed that pores in the oxide opened and closed during oscillations \cite{Aggour1995}. 
  
Most theoretical studies were motivated by the latter observation of opening and closing pores during oscillations. Many of them dealt with the question of how local, microscopically small oscillating domains synchronize to give rise to macroscopic oscillations in current density, not addressing the physico-chemical mechanism of how the local oscillations occur in the first place \cite{Ozanam1992, Grzanna2000a}. An exception hereof is the current-burst model suggested by Föll and coworkers. The model assumes that the ionic current through the oxide layer flows through pores that open at a certain strength of the electric field across the oxide layer, allowing for oxide growth. The pores close again only at a considerably lower field strength and only reopen when the oxide has been sufficiently etched such that the critical field strength for pore opening is reached again.
Monte-Carlo simulations of the current-burst model agree in many respects with experimentally observed oscillations. However, today we know that this picture is incomplete. 
The system exhibits birhythmicity, and thus, there are two distinct types of oscillation patterns, so-called high-amplitude oscillation (HAO) and low-amplitude oscillation (LAO) \cite{Tosolini2019, Schoenleber2012}. HAOs are observed under relatively high voltages and typically have a relaxation-type character, while LAOs are sinusoidally shaped and occur at lower voltages. Their coexistence in certain parameter intervals underlines that their oscillation mechanisms also differ. The current-burst model seems to describe HAOs.  

\begin{figure*}[htbp] 
\centering
\includegraphics[width=0.65\textwidth]{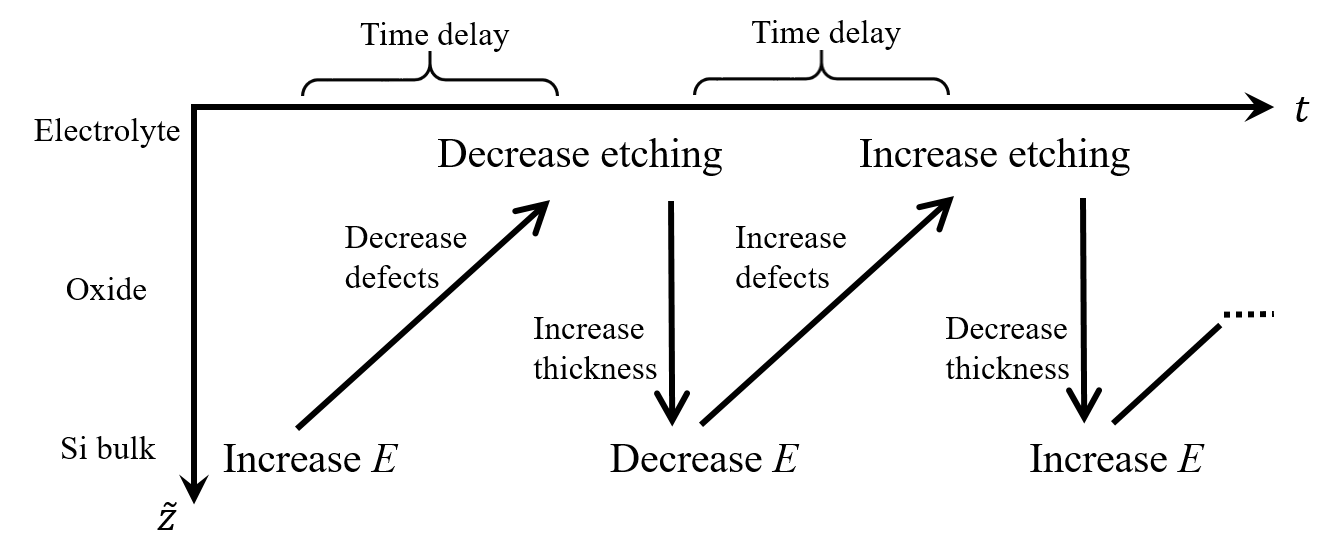}
\caption{A conceptual diagram of the silicon-oscillation mechanism. The vertical axis represents the coordinate perpendicular to the oxide layer, and the horizontal axis represents time. A change in electric field strength at the Si/oxide interface alters the average oxidation state of the Si oxide (the number of defects). After a time delay, the altered composition is exposed to the electrolyte interfaces, resulting in etching at a different rate. This, in turn, prompts a change in thickness and, consequently, a change in electric field strength that causes again an altered composition of the oxide. 
}
\label{fig:Schematic_mechanism}
\end{figure*}

In this article, we present a mathematical model for low-amplitude oscillations during Si electrodissolution that involves just a few electrochemical and chemical reaction steps, yet predicts the important dependencies on the experimental parameters. 
This basic physicochemical model predicts oscillations when the following three conditions are met: 
1. The dissolution rate of the oxide film increases with an increasing number of partially oxidized silicon species, below also called defects, in the oxide. 
2. The fraction of partially oxidized species decreases under a high electric field at the Si/oxide interface because they become fully oxidized. 
3. There is a sufficient time delay between the instant at which the defect is produced at the Si/oxide interface and the instant at which it is exposed to the electrolyte and thus etched. In other words, oscillations require a sufficiently thick oxide film. The possible importance of a delay for oscillations was first hypothesized by Smith and Collins, who assumed the formation of 'hard’ and ‘soft’ oxides that are etched at different rates \cite{Smith1992}. 
The interplay between etch rate and defect formation rate constitutes a negative feedback. However, the time delay suppresses the negative feedback for a certain time, causing an effective positive feedback that promotes stable oscillation. The fundamental oscillation mechanism is illustrated in Fig.~\ref{fig:Schematic_mechanism}. It does not seem to be specific to Si electrodissolution but may also apply to other semiconductor or metal electrodes in which anodic formation of oxide layers and their dissolution is accompanied by self-sustained oscillations.  

The paper is organized as follows: In section \ref{sec:Mathematical Model}, we derive the mathematical model that describes the dynamics of the oxide film. We first introduce the governing equations for the potential and the different Si-oxide species in the oxide layer (subsection \ref{section:GovEqChem}). 
In subsections \ref{section:BoundarySiOx} and \ref{section:boundaryOxEl}, we specify the boundary conditions for the Si-oxide species at the Si/oxide and the oxide/electrolyte interfaces, respectively. Then, we explain how we define a moving frame that accounts for the time-dependent motions of the two interfaces (subsection \ref{MovingFrame}), before we derive the boundary conditions for the electrostatic potential in subsection \ref{section:BCPotential}. This section is concluded by a discussion of the numerical implementation of the model (subsection \ref{section:NumImpl}).  In section \ref{sec:Results of the Full Model}, numerical simulations of the model are presented together with a comparison to experimental findings under potentiostatic (subsection \ref{section:potentiostatic}) and potentiodynamic (subsection \ref{section:Potentiodynamic}) conditions. A detailed analysis of the temporal evolutions of the various simulated quantities allows us then to deduce the oscillation mechanism in subsection \ref{section:OscilMech}. Based on the insights of the oscillation mechanism, we derive a simplified model that consists of  two coupled delay differential equations (section \ref{sec:simplifiedModel}). Simulation results of this simplified model together with a linear stability analysis are presented in section \ref{section:ResSimplModel}. Section \ref{section:Conclusion} concludes the article with a summary of the results and a short outlook. Experimental conditions and parameter values used in the simulations can be found in the Appendix.

\section{Mathematical Model} \label{sec:Mathematical Model}
The model calculates the temporal evolution of the state of the oxide layer bounded by two moving interfaces, the oxide/electrolyte and the Si/oxide interfaces.
For this purpose, we consider the following processes: charge-transfer reactions at the Si/oxide interface, diffusion and migration of defects within the oxide layer, and ion transfer as well as chemical etching reactions at the oxide/electrolyte interface.
In mathematical terms, we have to solve continuity equations for the chemical species in the oxide layer together with Poisson's equation, which determines the electrostatic potential within the oxide. 
The reactions at the two interfaces are part of the boundary conditions, as is the potentiostatic control of the experiment. 

The oxidation of Si at the Si/oxide interface occurs in many reaction steps and involves various  oxide species with Si oxidation states from I to IV. 
Furthermore, substoichiometric oxides may also be further oxidized within the oxide layer. 
For the sake of simplicity, we consider only three species in the model and neglect charge-transfer processes in the oxide. 
The species considered are fully oxidized silicon oxide, $\rm SiO_2$,  one substoichiometric oxide species, $\rm SiO$, and a charged defect in the oxide lattice, $\rm SiO^{2+}$, in other words, an $\rm O^{2-}$ vacancy.
These species are transported through the oxide and react further at the oxide/electrolyte interface.
 
\begin{figure}
\centering
\includegraphics[width=0.4\textwidth]{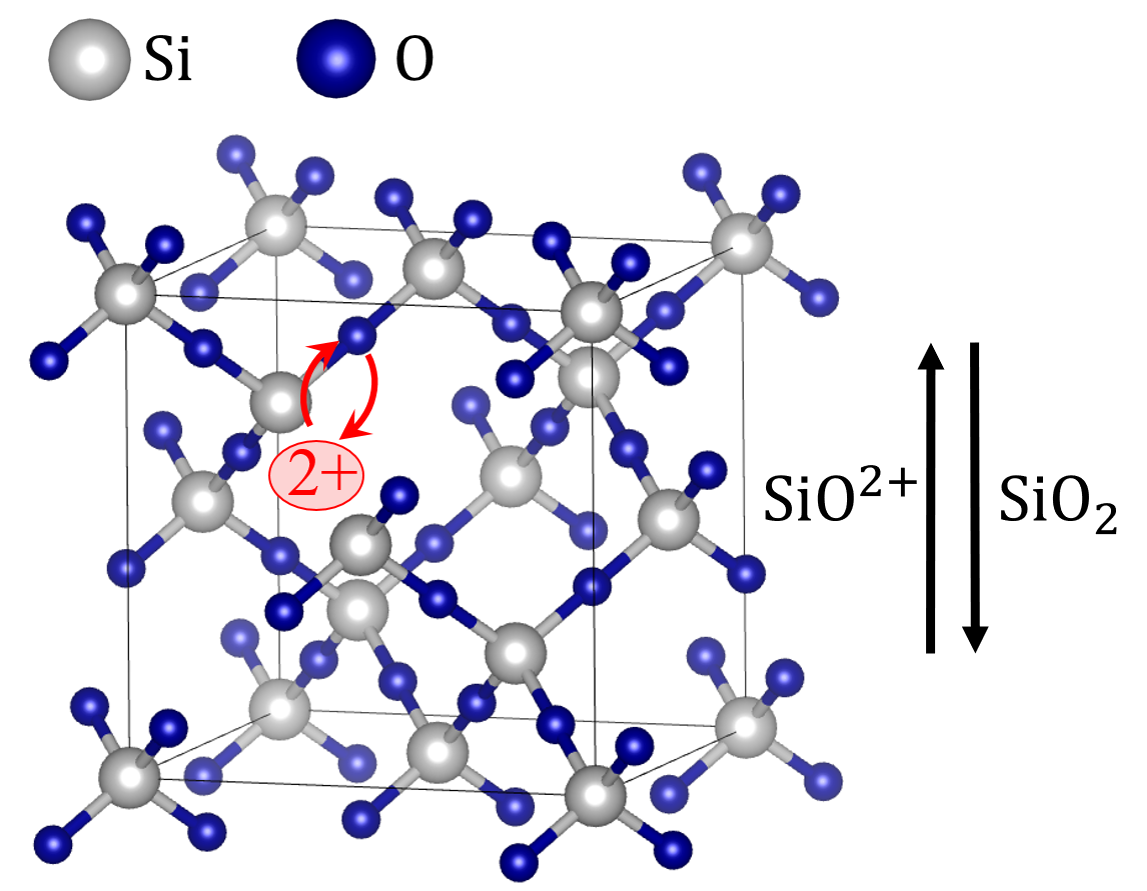}
\caption{A sketch of oxygen-vacancy transport through the $\rm SiO_2$ lattice.
}
\label{fig:SiO2_lattice}
\end{figure}

\begin{figure}
\centering
\includegraphics[width=0.49\textwidth]{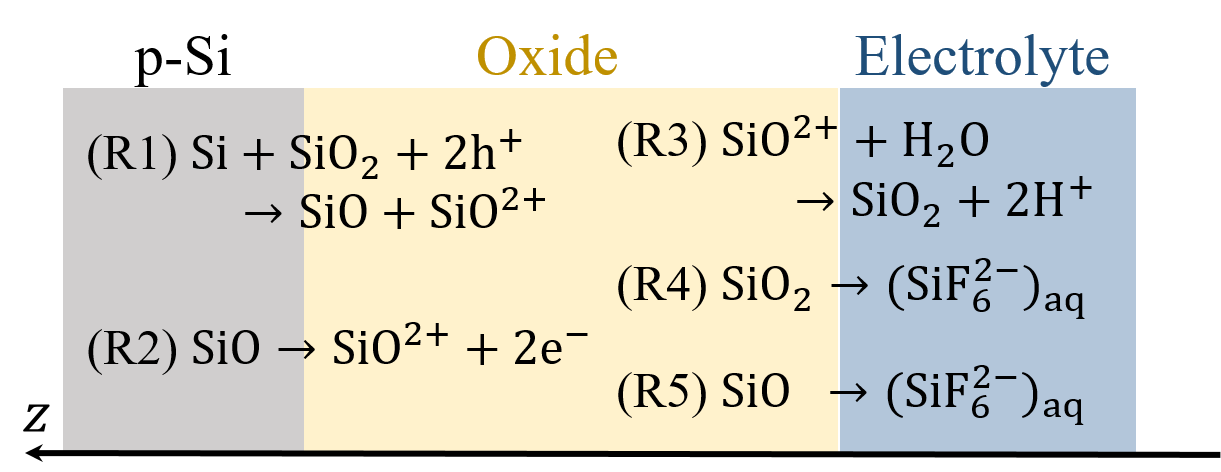}
\caption{A sketch of the considered reaction steps at the two interfaces.}
\label{fig:schematic_reactions}
\end{figure}

A positively charged oxygen vacancy moves by exchanging its position with neighboring $\rm O^{2-}$ ions, thereby carrying the current through the oxide and simultaneously leading to the buildup of $\rm SiO_2$ in the opposite direction. Owing to the electric field and the number density gradient of $\rm SiO^{2+}$, oxygen vacancies are transported toward the oxide/electrolyte interface, and thus $\rm SiO_2$ is transported toward the Si/oxide interface. This is depicted schematically in Fig.~\ref{fig:SiO2_lattice}.
Finally, at the oxide/electrolyte interface, oxygen vacancies are filled through rapid reaction with water and SiO and $\rm SiO_2$ are chemically etched. The processes are summarized in Fig.~\ref{fig:schematic_reactions}, and will be further motivated in the following subsections. The scheme can be seen as a type of point defect model, as originally developed for metal/oxide/electrolyte interfaces \cite{Macdonald_1992}.

In the following subsections we will derive the corresponding mathematical model.

\subsection{Governing Equations within the Si-Oxide layer} \label{section:GovEqChem} 
We assume that the oxide layer is laterally uniform and thus consider only variations along the spatial coordinate $z$ perpendicular to the oxide layer. The continuity equations for the three considered species read:
\begin{align}
\frac{\partial \rm [SiO^{2+}]}{\partial t}
=& D_{\rm SiO^{2+}}\frac{\partial^2\rm [SiO^{2+}]}{\partial z^2} \notag \\
&+\mu_{\rm SiO^{2+}}\frac{\partial}{\partial z}\left(\rm [SiO^{2+}]\; \frac{\partial \phi}{\partial z}\right) \label{eq:mass_cons_SiO2+} \\
\frac{\partial [\rm SiO]}{\partial t}
=& D_{\rm SiO}\frac{\partial^2[\rm SiO]}{\partial z^2}  \label{eq:mass_cons_SiO}  \\
[\rm SiO_2]=& \rho_{\rm atom} -\rm [SiO] - \rm [SiO^{2+}] \label{eq:mass_cons_SiO2}
\end{align}
The positive $z$ direction points from the electrolyte into the direction of the $\rm Si$ bulk.
The number density of species $\rm X_{\it i}$ is indicated by the square bracket notation [$\rm X_{\it i}$], and $\phi$ is the electrostatic potential. 
$D_{\rm X_{\it i}}$ is the diffusion coefficients of species $\rm X_{\it i}$ in the oxide, and $\mu_{\rm SiO^{2+}}$ the mobility of oxygen vacancies.
$\rho_{\rm atom}$ is the total $\rm Si$ atomic density of the bulk-silicon lattice.

The first term of eq.~(\ref{eq:mass_cons_SiO2+}) describes diffusion, and the second one migration of oxygen vacancies according to the Nernst-Planck approximation. As a neutral species, SiO is only subject to the diffusion eq.~(\ref{eq:mass_cons_SiO}). 
The number density of $\rm SiO_2$ follows the algebraic mass balance equation eq.~(\ref{eq:mass_cons_SiO2}).
For simplicity and clarity, the volume expansion that occurs when silicon is oxidized to silicon oxide is not taken into account.

The electrostatic potential $\phi$ obeys Poisson's equation:
\begin{align}
- \frac{\partial^2 \phi}{\partial z^2} &= \ \frac{2q {\rm [SiO^{2+}]}}{\epsilon_0 \epsilon_{\rm ox}}\label{eq:poisson}
\end{align}
where $q$, $\epsilon_0$, and $\epsilon_{\rm ox}$ are the elementary charge, the vacuum permittivity, and the relative permittivity of silicon oxide, respectively.
The relative permittivity of silicon oxide is assumed to be independent of its composition. 

In the following subsections, we specify the boundary conditions for the chemical species and the electrostatic potential at the two interfaces that describe $\rm Si$ electrodissolution at moderate voltages.

\subsection{Boundary Conditions for the Chemical Species at the Si/Oxide Interface} \label{section:BoundarySiOx}
At the Si/oxide interface, the oxide grows into the $\rm Si$ bulk by the generation of substoichiometric $\rm {Si}$ oxide through electric field-driven 'hopping' of oxygen ions from the $\rm Si$ oxide lattice into the space-charge layer of the Si, leaving positively charged oxygen vacancies behind.
The substoichiometric oxides are, in part, further oxidized. 
As a result, the  $\rm SiO_2$  lattice expands towards the silicon bulk, with a high number density of positively charged oxygen vacancies at the Si/oxide interface. 
In our minimal reaction scheme, we describe these processes by the following two reactions:
\begin{align}
\rm Si + SiO_2 + 2h^+ &\xrightarrow{} \rm SiO^{2+} + SiO \tag{R1} \label{eq:reaction1} \\
\rm SiO &\xrightarrow{} \rm SiO^{2+} + 2e^-. \tag{R2} \label{eq:reaction2}
\end{align}
It is known that the initial charge-transfer step involves valence-band holes, $\rm h^+$, while the further oxidation steps can occur through hole capture or electron injection \cite{Lewerenz1988}. 
Here, we consider an electron-injection step in (\ref{eq:reaction2}).

The kinetic equations for the rates of formation of $\rm SiO$ and oxygen vacancies according to (\ref{eq:reaction1}) and (\ref{eq:reaction2}), $r_1$ and $r_2$, respectively, are derived according to the following assumptions: The rate of the first oxidation step of a $\rm Si$ atom according to (\ref{eq:reaction1}) is proportional to the number density of $\rm SiO_2$ species and of valence-band holes at the Si/oxide interface, which is located at $z_{\rm b}$.
The number density of valence-band holes is proportional to the electric field at the interface, $E_{\rm inter}=$ \nicefrac{$\partial\phi$}{$\partial z$}$|_{z_{\rm b}}$ \cite{Sze_Ng_2006}. 
Furthermore, we assume that the transfer of oxide ions into the space-charge layer is an activated process that depends exponentially on $E_{\rm inter}$. 
We therefore express the oxidation rate $r_1$ as follows
\begin{align}
r_1 &= k_1 E_{\rm inter} {\rm [SiO_2]|_{\it z_{\rm b}}} \exp(\alpha_1 E_{\rm inter}) \label{eq:r1}
\end{align}
with the rate constant $k_1$ and the temperature dependent constant $\alpha_1$. 

\begin{figure}
\centering
\includegraphics[width=0.38\textwidth]{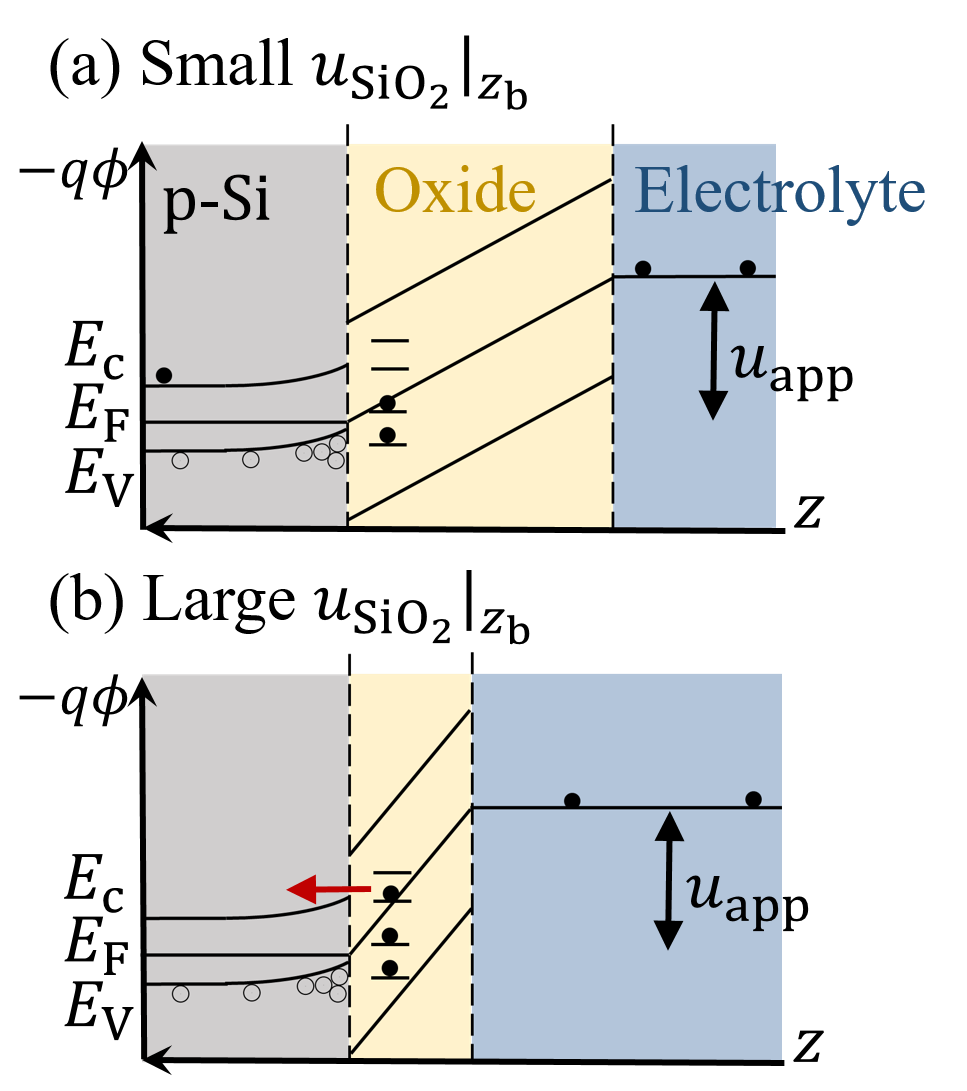}
\caption{
Schematic comparison of the degree of filling of surface states for a small $u_{\rm SiO_2}|_{\it z_{\rm b}}$ (a) and a large $u_{\rm SiO_2}|_{\it z_{\rm b}}$ (b).
The filled and unfilled dots represent electrons and holes, respectively.
For a small $u_{\rm SiO_2}|_{\it z_{\rm b}}$, the highest level of occupied surface states is lower than the conduction band energy of silicon bulk, and electron injection through the surface states is slow.
On the other hand, if $u_{\rm SiO_2}|_{\it z_{\rm b}}$ is large, the high energy of the occupied surface states causes (\ref{eq:reaction2}) to proceed rapidly.
}
\label{fig:sigmoid}
\end{figure}

The further oxidation of $\rm SiO$ through electron injection (\ref{eq:reaction2}) depends on the surface-state density and its occupation level. 
Electron injection through the surface states occurs only when the energy of occupied surface states reaches the bottom edge of the conduction band of the silicon bulk. 
The occupation level of a surface state increases as $-q\phi$ increases, which happens when the oxide becomes thinner and therefore the electric field at the interface becomes larger. 
We chose to express the occupation level of the surface states through the potential drop across the oxide monolayer adjacent to the Si bulk, $u_{\rm SiO_2}|_{\it z_{\rm b}}$.
 This is schematically depicted in Fig.~\ref{fig:sigmoid}.
Furthermore, we assume that the number of occupied surface states is limited towards higher energies, which we described by a sigmoidal dependence of the oxidation rate $r_2$ on $u_{\mathrm {SiO_2}|_{z_{\rm b}}}$. 
We therefore approximate  the rate $r_2$ by
\begin{align}
r_2 &= \frac{k_2 {\rm [SiO]|_{\it z_{\rm b}}} } {1 + \exp(\alpha_2 (\alpha_3 - u_{\rm SiO_2|_{\it z_{\rm b}}}))}, \label{eq:r2}
\end{align}

\noindent where $k_2$ is a rate constant and the constants $\alpha_2, \alpha_3$ determine the shape of the sigmoidal function.
Here, $r_1$ and $r_2$ are defined so that their dimension is $\rm 1/(m^2s)$.

The boundary conditions for the chemical variables $[\rm SiO]$ and $[\rm SiO^{2+}]$ at the Si/oxide interface, i.e. at $z\!=\!z_{\rm b}$ are given by the fluxes of these species through the interface, which result from the reaction rates eq.~(\ref{eq:r1}) and eq.~
(\ref{eq:r2}).
\begin{align}
\left. \frac{\partial [\rm SiO^{2+}]}{\partial z} \right|_{{\it z_{\rm b}}} &= -\frac{r_1+r_2}{D_{\rm SiO^{2+}}} \\
\left. \frac{\partial [\rm SiO]}{\partial z} \right|_{{\it z_{\rm b}}} &= -\frac{r_1-r_2}{D_{\rm SiO}}
\end{align}

\subsection{Boundary Conditions for the Chemical Species at the Oxide/Electrolyte Interface}\label{section:boundaryOxEl}
At the oxide/electrolyte interface, the oxygen vacancies are filled through the reaction with water according to (\ref{eq:ion-iontransfer}). 
In this way, oxygen ions enter the oxide and $\rm SiO_2$ is formed.   
\begin{align}
\rm SiO^{2+} +H_2O  &\xrightarrow{\rm fast} \rm SiO_2 +2H^+\tag{R3}
\label{eq:ion-iontransfer}
\end{align}
The overall electrooxidation reaction of $\rm Si$, which is usually written as (\ref{eq:ox}) 
thus consists of charge-transfer reactions at the Si/oxide interface and the ion-transfer reaction at the oxide/electrolyte interface (in our case (\ref{eq:ox})=(\ref{eq:reaction1})+(\ref{eq:reaction2})+2 $\times$ (\ref{eq:ion-iontransfer}) with $\lambda_{\rm VB}=2$).

The ion-transfer reaction (\ref{eq:ion-iontransfer}) is assumed to be so fast that the number density of oxygen vacancies in the oxide layer adjacent to the electrolyte, hereafter after the top oxide layer, described with the subscript '$\it z_{\rm t}$', is assumed to be 0:
\begin{align}
\rm  [SiO^{2+}]|_{\it z_{\rm t}}  & = 0 \label{eq:SiO2+_top} 
\end{align}
The boundary condition for [\rm SiO] is easiest expressed in a moving frame (see subsection \ref{MovingFrame}). 
We denote the  spatial coordinate in the moving frame by $\tilde{z}$. 
As will become clear below, in the moving frame, the flux of SiO species at the oxide/electrolyte interface is zero. 
The boundary condition for the number density of [SiO] thus reads:
\begin{align}
\left. \frac{\partial [\rm SiO]}{\partial z} \right|_{{\it z_{\rm t}}} &= 0
\end{align}

\subsection{Interface Motion and Moving Frame \label{MovingFrame}}
So far, we have specified the boundary conditions at the interfaces Si/oxide and oxide/electrolyte, or, expressed more formally, at $z=z_{\rm b}$ and $z=z_{\rm t}$. However, both interfaces, and consequently the positions of $z_{\rm b}$ and $z_{\rm t}$, move in time, which makes solving the model numerically very complicated. 
This challenge can be overcome when using a moving frame. 

The velocity $v_{\rm ox}$ with which the Si/oxide interface moves towards the Si-bulk (in the positive z-direction in the laboratory frame) is determined by the rate with which SiO is formed (\ref{eq:reaction1}), and thus reads
\begin{align}
v_{\rm ox} &= \frac{r_1}{\rho_{\rm atom}}.
\label{eq:oxidation_}
\end{align}

At the oxide/electrolyte interface, the $\rm Si$ oxide is etched chemically by fluoride species from the electrolyte. 
We do not formulate the stoichiometry of individual etching steps here since they are irrelevant for the model.
Rather, we assume that substoichiometric $\rm Si$ oxides are etched faster than $\rm SiO_2$ \cite{Smith1992}
\begin{align}
\mathrm {SiO_2}|_{z_{\rm t}}\xrightarrow [\rm HF, HF_2^-]{k^{\rm SiO_2}_{{\rm etch}}}\rm (SiF_6^{2-})_{aq} \tag{R4} \label{eq:SiO2_aq}\\
\mathrm {SiO}|_{z_{\rm t}} \xrightarrow [\rm HF, HF_2^-]{k^{\rm SiO}_{{\rm etch}}\;>\;k^{\rm SiO_2}_{{\rm etch}}}\rm (SiF_6^{2-})_{aq}. \tag{R5} \label{eq:SiO_aq}
\end{align} 
Here, the $k^{\rm i}_{\rm etch}$ are rate constants, and $\rm (SiF_6^{2-})_{aq}$ indicates silicon 
hexafluoride ions dissolved in the electrolyte. 
Note that according to eq.~(\ref{eq:SiO2+_top}), we do not have to take into account etching of $\rm SiO^{2+}$ species.  
Thus, the overall etch rate, $v_{\rm etch}$, measured in m/s, can be written as
\begin{align}
    v_{\rm etch} &= k_{v_{\rm etch}}({\rm [SiO_2]|_{\it z_{\rm t}}} +k_{\rm SiO}\rm [SiO]|_{\it z_{\rm t}}) \label{eq:etching}
\end{align}
$k_{v_{\rm etch}}$ is proportional to $k_{\rm etch}^{\rm SiO_2}$.
$k_{\rm SiO}>1$ defines how much faster SiO is etched than $\rm SiO_2$. 

\begin{figure}
\centering
\includegraphics[width=0.5\textwidth]{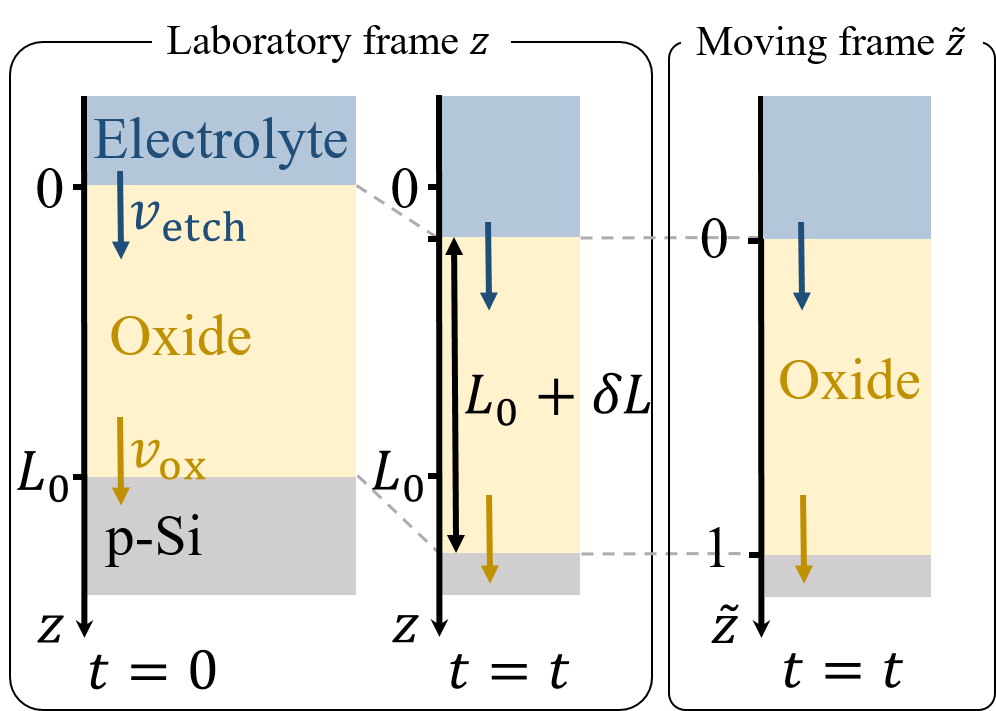}
\caption{
Definitions of laboratory frame $z$ (left) and moving frame $\tilde{z}$ (right).
}
\label{fig:moving_frame}
\end{figure}

We define the moving coordinate $\tilde{z}$ so that, independent of time, the oxide/electrolyte interface is always at $\tilde{z}=0$ and the Si/oxide interface at $\tilde{z}=1$, as illustrated in Fig.~\ref{fig:moving_frame}.
Note that $\tilde{z}$ is a unit-less quantity.
Hence, as long as we calculate the physical quantities in this moving frame, the position of the two boundaries of the system is fixed.
If we denote the initial oxide thickness at $t=0$ by $L_0$, we can write for the oxide thickness at $t=t$ 
\begin{align}
L(t) = L_0 + \delta L(t),
\end{align}
where
\begin{align}
\delta L(t) = \int^t_0 ( v_{\rm ox}(t')- v_{\rm etch}(t') ) {\rm d}t'.
\end{align}
Analogously, we define the oxide/electrolyte interface displacement, $\delta L_{\rm etch}(t)$ between $t=0$ and $t=t$ by
\begin{align}
\delta L_{\rm etch}(t) = \int^t_0 v_{\rm etch}(t') {\rm d}t'. 
\end{align}
The conversion between $z$ and $\tilde{z}$ can then be written as
\begin{equation}
\tilde{z} = \frac{1}{L(t)}\left(z-\delta L_{\rm etch}(t)\right). \label{eq:z_relation_math}
\end{equation}
Here, $\delta L_{\rm etch}(t)$ represents the offset of the two frames, and $1/L(t)$ defines the stretching ratio of the two different coordinates.

In the moving coordinate system, the time evolution of a number density at each position $\tilde{z}$ is caused by the local time evolution in the laboratory frame and the moving frame's apparent effect.
The relation between the two different time derivatives is expressed as follows
\begin{align}
\frac{\rm D}{\rm D \it t} &=\frac{\partial}{\partial t} - \frac{\partial \tilde{z}}{\partial t}\frac{\partial}{\partial \tilde{z}}.
\end{align}
Here, $\partial / \partial t$ represents the derivative at the static (laboratory) $z$ coordinate, and $\rm D / D \it t$ that at the $\tilde{z}$ coordinate. 
It takes into account how a certain quantity is associated with changes over time due to the motion of the coordinate system. 
$\partial \tilde{z} / \partial t$  is given by
\begin{align}
\frac{\partial \tilde{z}}{\partial t} 
= - \frac{1}{L(t)}\left( (v_{\rm ox}(t) -v_{\rm etch}(t)) \tilde{z}+v_{\rm etch}(t)\right), \label{eq:dz_dt}
\end{align}
whereby the first term proportional to $\tilde{z}$ takes into account that the oxide-layer thickness changes with time, and the second one is the velocity with which the oxide/electrolyte interface moves, which also depends on time. 

\subsection{Boundary Conditions for the Electrostatic Potential}\label{section:BCPotential}
\begin{figure}
\centering
\includegraphics[width=0.25\textwidth]{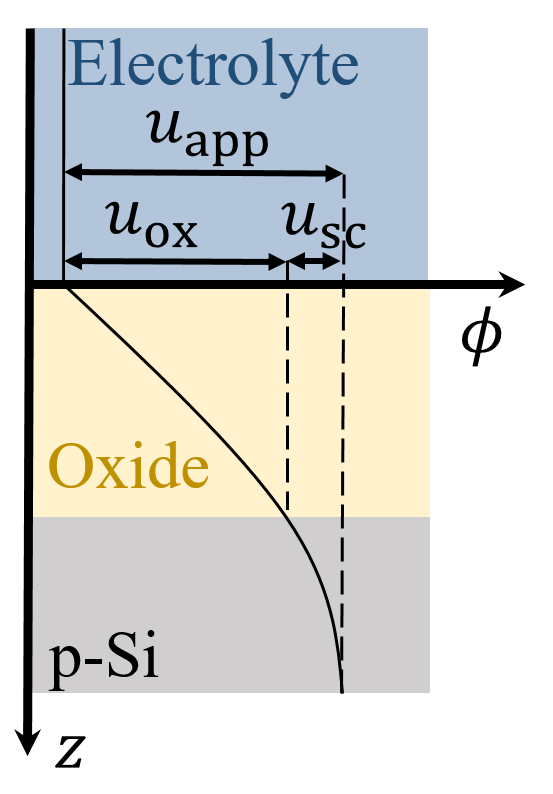}
\caption{
Schematic representation of the potential distribution across the electrode/electrolyte interface.
}
\label{fig:u_eff}
\end{figure}

To complete the model and to be able to solve Poisson's eq.~(\ref{eq:poisson}), we need to specify the boundary conditions of the electrostatic potential. 
A schematic potential distribution across the Si/oxide/electrolyte layers is shown in Fig.~\ref{fig:u_eff}. 
The potential drops across the space-charge layer, $u_{\rm sc}$, and the total oxide layer, $u_{\rm ox}$, are much larger than those across the silicon bulk and the electrolyte (including the double layer). Therefore, we neglect the latter two and assume that the applied potential, $u_{\rm app}$, is distributed over the oxide layer and the space-charge layer
\begin{align}
u_{\rm app} = u_{\rm ox} + u_{\rm sc}. \label{eq:u_app}
\end{align}
Choosing the oxide/electrolyte interface as the origin of the potential scale, the boundary conditions at the oxide/electrolyte and Si/oxide interfaces are given by
\begin{align}
\phi_{\it z_{\rm t}} &= 0 \label{eq:phi_top}, \\
\phi_{\it z_{\rm b}} &= u_{\rm app} - u_{\rm sc}, \label{eq:phi_bottom}
\end{align}
respectively.

\begin{figure}
\centering
\includegraphics[width=0.35\textwidth]{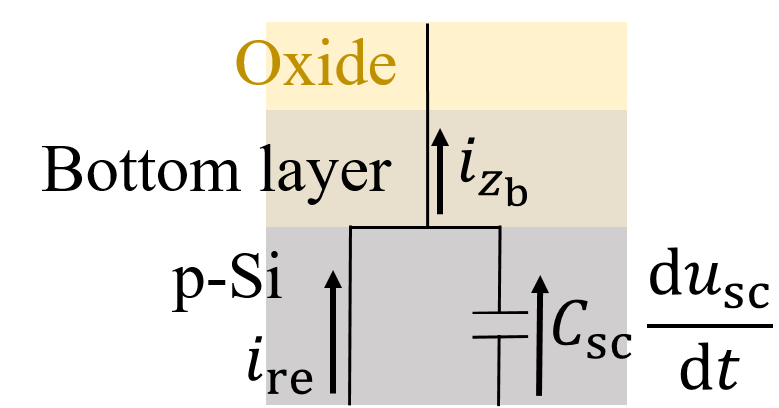}
\caption{
The equivalent circuit of the Si/oxide interface.
}
\label{fig:Kirchhoff}
\end{figure}

Next, we determine the temporal evolution of $u_{\rm sc}$. Therefore, we model the Si/oxide interface by the equivalent circuit depicted in Fig.~\ref{fig:Kirchhoff}.
According to Kirchhoff's current law, 
\begin{align}
i_{\it z_{\rm b}} =& i_{\rm re} + C_{\rm sc} \frac{{\rm d}u_{\rm sc}}{{\rm d}t},
\label{eq:Kirchhoff}
\end{align}
where $i_{\it z_{\rm b}}$, $i_{\rm re}$, and $C_{\rm sc}$ are the migration current density into the oxide, the electrochemical reaction current density, and the capacitance per unit area at the space-charge layer, respectively.
The migration current density at the interface is given by
\begin{equation}
i_{\it z_{\rm b}}  = \sigma_{\it z_{\rm b}} E_{\it z_{\rm b}},
\end{equation}
where $\sigma_{\it z_{\rm b}}$ and $E_{\it z_{\rm b}}$ are the conductivity and the electric field at the bottom of the oxide, $z=z_{\rm b}$. 
Note that we here take the derivative when approaching the boundary from the oxide side, i.e., from the $+z$ direction, whereas the electric field at the boundary approached from the Si side (the $-z$ direction) is denoted by $E_{\rm inter}$.
The resistivity of the oxide at $z_{\rm b}$ is calculated according to
\begin{align}
\sigma_{\it z_{\rm b}} &= 2 q ([{\rm SiO^{2+}}]|_{\it z_{\rm b}} + k_{\rm leak}) \mu_{\rm SiO^{2+}}. 
\end{align}
The small constant $k_{\rm leak}$ is added to avoid numerical difficulties.
The reaction current density, $i_{\rm re}$ is defined by
\begin{equation}
i_{\rm re}   = 2q (r_1 + r_2)   \label{eq:i}
\end{equation}

The time evolution of $u_{\rm sc}$ is calculated using eq.~(\ref{eq:Kirchhoff}).

Finally, the electric field at the space-charge layer, $E_{\rm inter}$, which is necessary to calculate $r_1 $ and $ r_2$ is a function of $u_{\rm sc}$ and given by
\begin{align}
E_{\rm inter} &= \frac{\sqrt{2} k_{\rm B}T} {q L_{\rm D}} F_{\phi}( u_{\rm sc} ) \\ 
F_{\phi}(u_{\rm sc} ) &= \sqrt{\exp \left(- \frac{q}{k_{\rm B}T} u_{\rm sc} \right) + \frac{q}{k_{\rm B}T} u_{\rm sc} -1},
\end{align}
where $k_{\rm B}$, $T$, and $L_{\rm D}$ are the Boltzmann constant, temperature, and Debye length, respectively.

This relation is calculated by approximating the Si/oxide/electrolyte interface as a semiconductor-insulator-conductor connection \cite{Sze_Ng_2006}.

\subsection{Numerical Implementation}\label{section:NumImpl}
\begin{figure}[htbp] 
\centering
\includegraphics[width=0.2\textwidth]{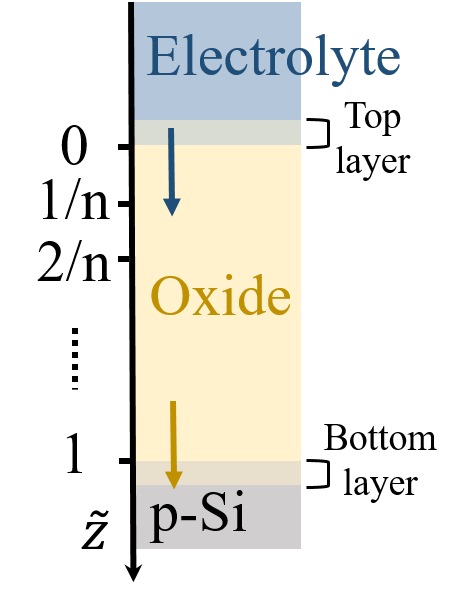}
\caption{
Discretization of the oxide. 
}
\label{fig:discretization}
\end{figure}

An illustration of our spatial discretization scheme is shown in Fig.~\ref{fig:discretization}.
We discretized the oxide into $n = 25$ grid points.
Because the thickness of the oxide is fixed at 1 on the $\tilde{z}$ coordinate, the physical thickness of each discretized length element, $l_{\rm ox}$, changes over time.
For easier implementation of the boundary conditions, we define additional top and bottom layers which sandwich the discretized oxide and have a fixed length, $l_{\rm m}$. 

\begin{equation}
l_{\rm ox}|_{\it z_{\rm t}} = l_{\rm ox}|_{\it z_{\rm b}} = l_{\rm m} = {\rm const}
\end{equation}
Numerical simulations were conducted using MATLAB R2024b.
Time integration was performed using the explicit Euler method using a time step $\Delta t = 1\times10^{-5}$.
All parameters used in the model are summarized in the appendix \ref{section:Parameters of the Full Model}.

\section{Results of the Full Model} \label{sec:Results of the Full Model}
In this section, the numerical simulation results of the complete model are presented and
compared with the experimental results.
\subsection{Potentiostatic Conditions}\label{section:potentiostatic}

\begin{figure}[htbp] 
\centering
\includegraphics[width=0.43\textwidth]{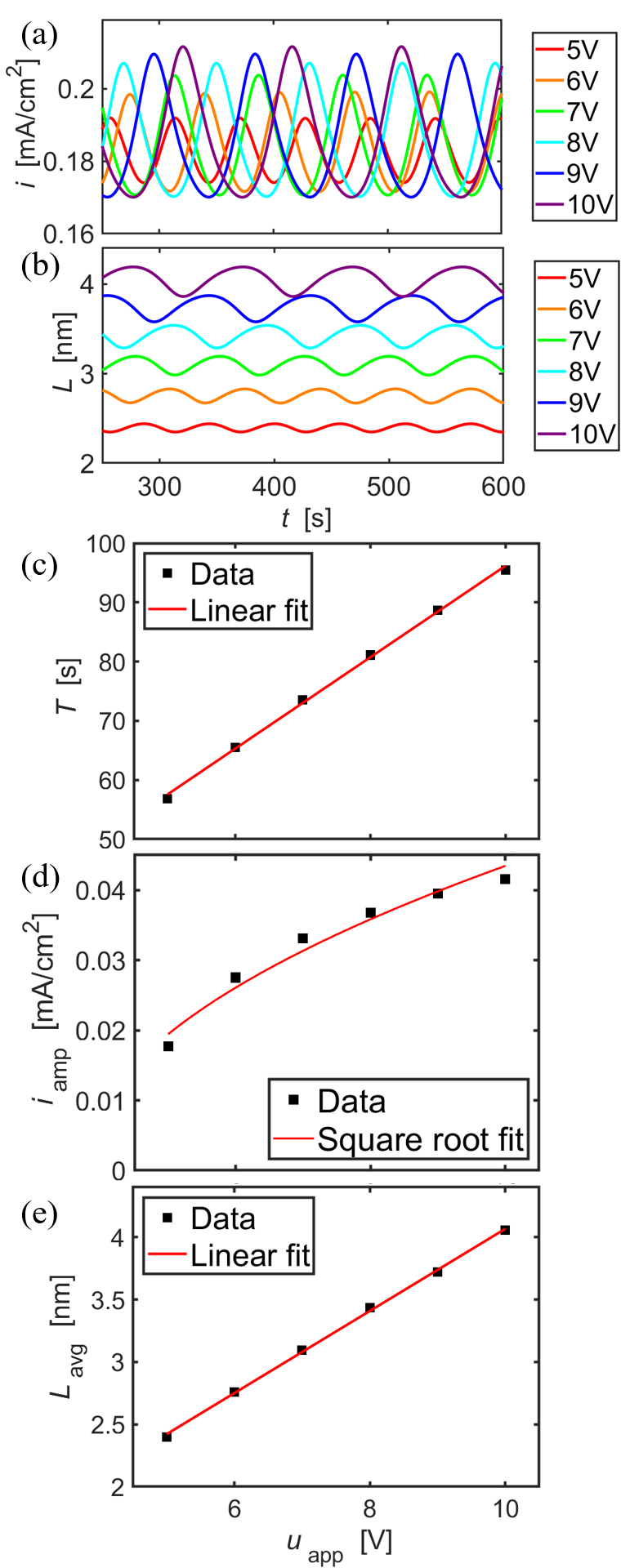}
\caption{
Simulation results of the full model. (a) current time series, (b) time series of oxide-layer thickness, (c) oscillation period, (d) current amplitude, and (e) average oxide-layer thickness vs. applied voltage.
}
\label{fig:multi_current_thickness}
\end{figure}

The model exhibits self-sustained current and oxide-layer thickness oscillations under potentiostatic conditions, i.e., at a constant value of the applied voltage $u_{\rm app}$.
More precisely, with the parameter values used, oscillations were obtained for $u_{\rm app}\ge4.6~\rm V$ while for  $u_{\rm app} < 4.6~\rm V$, the etching rate exceeded the oxidation rate and no oxide layer formed.

Simulated time series of the current density (a) and the oxide-layer thickness (b) at various values of $u_{\rm app}$ in the oscillatory region are shown in Fig.~\ref{fig:multi_current_thickness}. 
The oscillation amplitudes and periods obviously change with the applied voltage. 
Furthermore, the average oxide-layer thickness increases with $u_{\rm app}$, while the mean current density changes only marginally. 

The dependence of the oscillation period, the amplitude of the current oscillations, and the time-averaged oxide depth on $u_{\rm app}$ is shown in plates (c), (d), and (e).
The period of the oscillation increases linearly with $u_{\rm app}$ while the current amplitude follows a square root relationship.
The nearly linear increase of the mean oxide-layer thickness with the applied potential at an almost constant average current density reflects the fact that the time-averaged electric field strength adjusts to a similar value regardless of the applied potential in a stable oscillatory state. 
This picture agrees with the ideas for film growth in the point defect model \cite{Macdonald_1992}.

\begin{figure}
\centering
\includegraphics[width=0.43\textwidth]{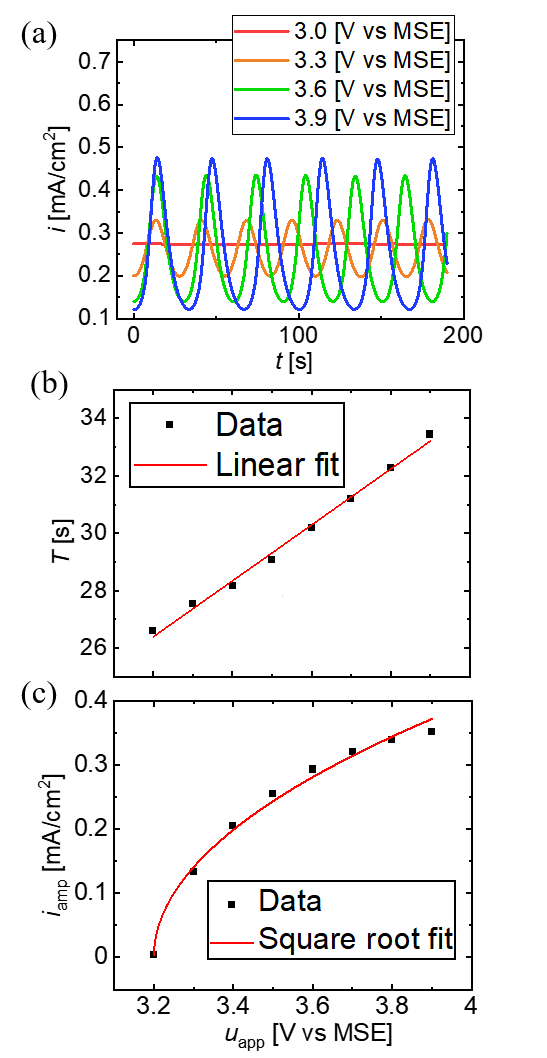}
\caption{
Experimental results. (a) Current time series for various values of the applied voltage, (b) oscillation period, (c) current amplitude vs. applied voltage. In all experiments, a series of external resistance of 1.77 $\rm k \Omega cm^2$ was inserted between the working electrode and the potentiostat.
}
\label{fig:Experiment_potentiostatic}
\end{figure}
For comparison, experimental results are shown in Fig.~\ref{fig:Experiment_potentiostatic}. 
The measurements were conducted under potentiostatic conditions in the range between 3 to 4 V vs. MSE with a series external resistance of 1.77 $\rm k \Omega cm^2$.
Further details of the experiments are summarized in the Appendix \ref{sec:Experimental Conditions}.
Obviously, all the characteristics of the oscillations from the simulations agree with the experimental observations. 
In both cases, the shape of the current oscillations is nearly harmonic while the mean current hardly changes with the applied voltage (Fig.~\ref{fig:Experiment_potentiostatic} (a)). Furthermore, the oscillation period (Fig.~\ref{fig:Experiment_potentiostatic} (b)) and the amplitudes of the current (Fig.~\ref{fig:Experiment_potentiostatic} (c)) increase linearly and square root-like with $u_{\rm app}$, respectively. 
In addition, current oscillations only occur  when the applied potential exceeds a threshold value of 3.2 V vs MSE, and the average current level stays nearly the same.
Finally, the linear increase of the average oxide thickness with the applied voltage agrees with the experimental result presented in \cite{Duportal_2024}.

\subsection{Potentiodynamic Conditions}\label{section:Potentiodynamic}

\begin{figure}
\centering
\includegraphics[width=0.45\textwidth]{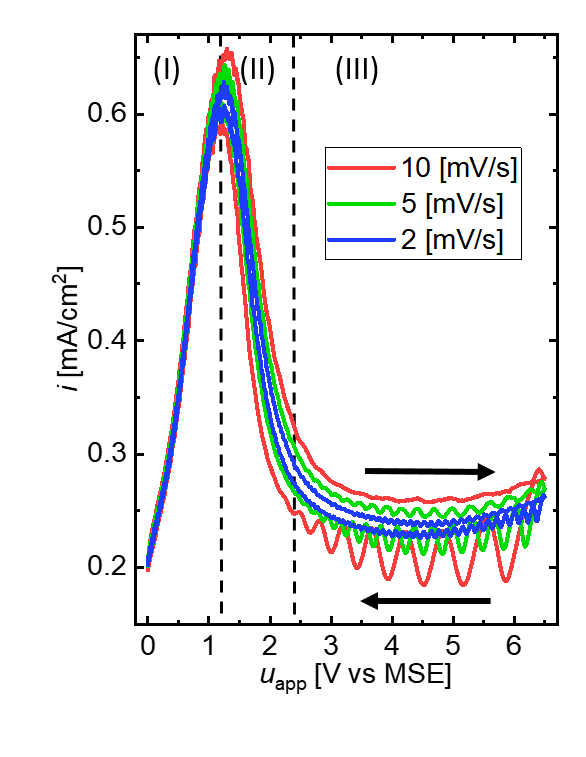}
\caption{
Experimental current voltage curves with different scan speeds of 10 mV/s (red), 5 mV/s (green), and 2 mV/s (blue). 
The experimental details are given in the Appendix \ref{sec:Experimental Conditions}.
}
\label{fig:CV_diff_speed}
\end{figure}

Having seen that the simulations at fixed parameter values agree well with the experimental results, we next compare experiments and simulations at cyclic voltage sweeps. Typical experimental current-voltage curves obtained at different scan speeds are depicted in Fig.~\ref{fig:CV_diff_speed}.
They can be roughly divided into three regions. 
Negative to the current maximum at $u_{\rm app}$=1.3 V vs MSE is the electropolishing region (I) in which a surface oxide forms and dissolves at equal rates such that no stable oxide builds up on the Si surface \cite{Zhang2001}. 
In the negative differential resistance region (II) between 1.3 and 2.5 V vs MSE, a stable oxide film develops on the electrode surface. At the lower potential edge of that region, the oxide composition is dominated by suboxides. They are increasingly oxidized to SiO$_2$ with increasing voltage \cite{Salman2019}. 
Region (III) in the voltage interval between 2.5 and 6.5 V vs MSE is the oscillatory region that we describe in our model. 
Here, the current density is limited by the etching speed, and sustained current oscillations are observed at any constant value of $u_{\rm app}$.
However, when the potential is swept, in the forward scan towards higher potentials, the oscillations are suppressed or have a considerably smaller amplitude than in the backward scan. 
Moreover, the average current density is higher in the forward scan than in the backward scan. 
This hysteresis increases with the scan speed.

\begin{figure}
\centering
\includegraphics[width=0.5\textwidth]{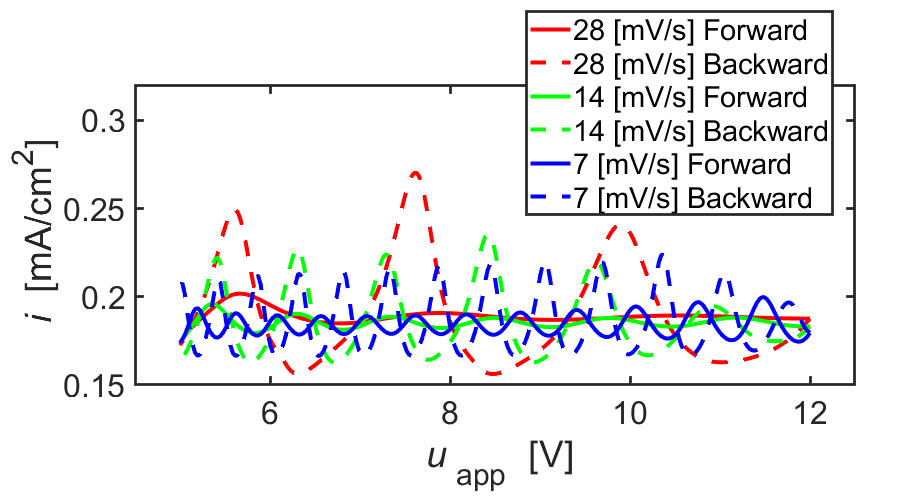}
\caption{
Simulated current potential curves with different scan speeds of 28 mV/s (red), 14 mV/s (green), and 7 mV/s (blue). 
The forward scans are shown in solid lines, and the backward scans in dashed lines.
}
\label{fig:multi_CVs}
\end{figure}

Simulations of current-voltage curves during cyclic voltage sweeps in region (III) covered by the model are shown for different scan rates in Fig.~\ref{fig:multi_CVs}.
The forward scans are shown as solid lines, and the backward scans as dashed lines.
At all scan rates, the oscillation amplitudes are larger for the backward scan than for the forward scan.
Furthermore, comparing the oscillation amplitudes at the different scan rates, they increase as the scan rate increases. 
These features show good agreement with the measurements.
However, different from the experimental data, a hysteresis in the average current density between the forward and reverse scan is not clearly recognizable in the simulated curves. Nevertheless, the qualitative agreement of the trends of the oscillation amplitudes with both scan rate and scan direction can be regarded as a validation of the model. 

\subsection{The Interpretation of the Oscillation Mechanism}\label{section:OscilMech}
\begin{figure}
\centering
\includegraphics[width=0.45\textwidth]{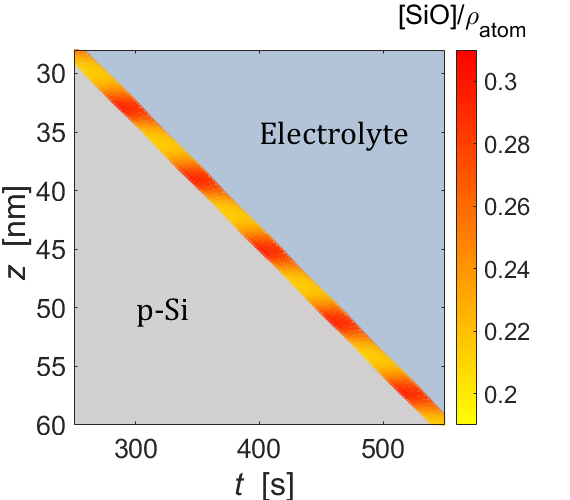}
\caption{
Color mapping of the spatio-temporal evolution of the  $\rm SiO$ fraction (normalized to the number density of Si) at $u_{\rm app} = 5$ V.
The upper right and lower left triangular areas are filled with electrolyte and p-Si bulk, respectively.
}
\label{fig:diago_SiO}
\end{figure}

So far, we have discussed simulation results that can be compared to experimental observations. 
In this subsection, we will look at the temporal evolution of simulated quantities that provide further insight into the oscillation mechanism. 
Fig.~\ref{fig:diago_SiO} depicts the spatio-temporal evolution of [SiO] in the oxide layer during oscillations, where the spatial coordinate $z$ is the static laboratory coordinate. 
The blue upper and gray lower triangles show where the electrolyte and the Si bulk are located.
It can be seen that the growth of the oxide layer into the Si bulk and the etching of the oxide by the electrolyte occur with a nearly constant velocity while the space-time plot of the [SiO] forms approximately a horizontal stripe pattern. 
The latter indicates that transport of SiO through diffusion is negligible. 
Rather, the suboxide defects remain very close to the $z$ position of their formation and only reach the electrolyte interface as the overlying oxide is etched. 
We denote the time delay between formation and etching of oxide species by $\tau$. 
In other words, $\tau$ is the time required to etch the oxide layer once, and thus, $\tau$ is proportional to the oxide thickness. 

\begin{figure}
\centering
\includegraphics[width=0.5\textwidth]{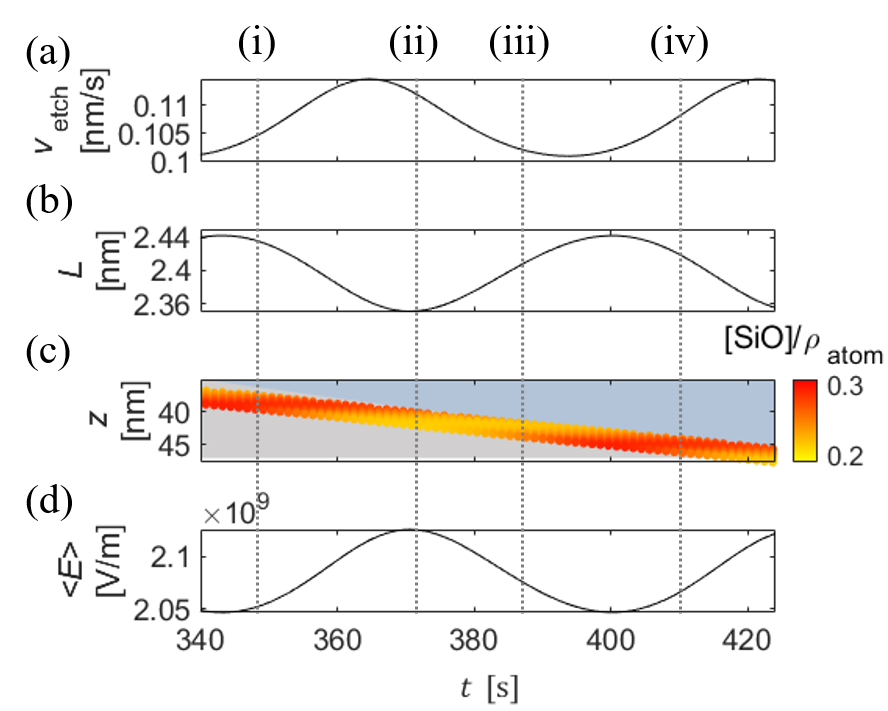}
\caption{
Simulated time series of (a) etching speed, (b) oxide-layer thickness, (c) fraction of partially oxidized defects, and (d) averaged electric field intensity across the oxide at $u_{\rm app}=5$ V.
}
\label{fig:four_data_labels}
\end{figure}

The role that the time delay, $\tau$ plays in the oscillation mechanism becomes evident when zooming in on the development of the oxide layer over slightly more than one oscillation period and analyzing the minor changes in etching rate and oxide-layer thickness in more detail. 
Fig.~\ref{fig:four_data_labels} displays the section of Fig.~\ref{fig:diago_SiO} between 330 and 414 s in plate (c) together with the etching velocity of the oxide, the oxide-layer thickness and the temporal evolution of average electric field strength across the oxide in plates (a), (b), and (d), respectively. 
The oscillation mechanism can be deduced from the changes at the four points in time, which are labeled (i)-(iv). 

At time (i), the electric field just starts to increase, which causes the reaction rate of (\ref{eq:reaction2}) to increase more strongly than that of (\ref{eq:reaction1}). 
In other words, the relative concentration of SiO at the bottom layer of the oxide decreases, as evidenced by the transformation from red to orange at the bottom of the oxide layer in plate (c). 
(ii) marks the time when this lowered SiO defect number density reaches the oxide/electrolyte interface. 
Hence, time (ii) is $\tau$ later than time (i).
The lower defect number density causes the etch rate to decrease such that the thickness of the oxide increases, which, in turn, reduces the electric field intensity. 
At (iii), the electric field intensity is so low that the SiO number density increases again.
$2\,\tau$ later than (i), at (iv), this increased number density of SiO reaches the oxide/electrolyte interface and causes the etching velocity to increase again, resulting in a decrease of the oxide thickness and an increase of the electric field, and so the cycle starts anew.
These results validate the conjectured mechanism  summarized in Fig.~\ref{fig:Schematic_mechanism}. 

The oscillation mechanism entails that approximately two oxide layers are etched during one oscillation period, which was also found experimentally \cite{Duportal_2024}.
Furthermore, it explains why oscillations occur only above a threshold voltage and why the period of the oscillation increases with increasing potential.
Also these two model predictions agree well with experimental results \cite{Duportal_2024}, see also Fig.~\ref{fig:Experiment_potentiostatic}.

The above considerations revealed that the time delay $\tau$ between the change in the ratio of the SiO and $\rm SiO_2$ number densities at the Si/oxide interface and its effect on the etch rate resumes the role of a positive feedback and is essential for the occurrence of the oscillations. 
The relations between (i) and (iii) ( $E\nearrow$ $\rightarrow$ $\rm [SiO]\searrow$ $\rightarrow$ $v_{\rm etch} \searrow$ $\rightarrow$ $L\nearrow$ $\rightarrow$ $E \searrow$) as well as between (ii) and (iv) form negative feedback loops that damp perturbations. 
Only when the delay is sufficiently large, i.e., the oxide sufficiently thick, can oscillations occur. 
On the other hand, any specific reaction or transport steps within the oxide seem to be irrelevant for the oscillation mechanism. 
This suggests that the model can be considerably simplified, which we will do in the next section.

\section{Simplified Model} \label{sec:simplifiedModel}
As discussed below, the oxide layer takes a passive role in the oscillation mechanism. 
Its only role is to provide a time delay between the formation and the etching of Si-oxide species. 
In this section, we show that indeed the electrochemical oscillation can be qualitatively described using a time-delay model with only two variables: the oxide-layer thickness, $l$ and the number density of substoichiometric oxide species, $\rm [SiO]$.
With this simplified model, we only aim to capture the qualitative behavior. All variables and parameters are dimensionless. 

\subsection{Oxide-Layer Thickness $l$}
The time evolution of the oxide-layer thickness, $l$ is determined by the difference in the rates with which bulk Si is oxidized, i.e. the velocity with which the Si/oxide interface moves, $v_{\rm ox}$, and the etching rate, that determines the velocity of the oxide/electrolyte interface, $v_{\rm etch}$.
\begin{align}
\frac{{\rm d} l}{{\rm d}t} &= v_{\rm ox} -v_{\rm etch} \label{eq:v_diff}.
\end{align}
As above, we define the velocity of the Si/oxide interface by the reaction speed of (\ref{eq:reaction1}).
When $\rm Si$ and $\rm SiO_2$ are present in large quantities at the Si/oxide interface, the reaction rate is determined by the number of valence-band holes at the interface, which depends on the electric field at the interface.
For the sake of simplicity, we assume here that $v_{\rm ox}$ is proportional to the electric field, $E(t)$ and that the electric field at the interface can be approximated by the average field across the oxide layer
\begin{align}
v_{\rm ox}(t) &= \gamma_1 E(t) \label{eq:v_ox},
\end{align}
where $\gamma_1$ is a constant. 
Furthermore, since the potential drop across the space-charge layer is negligibly small compared to the one across the oxide layer, we approximate $E(t)$ by 
\begin{align}
E(t) &=\frac{u_{\rm app}}{l(t)}. \label{eq:E=U/l}
\end{align}
\begin{figure}
\centering
\includegraphics[width=0.42\textwidth]{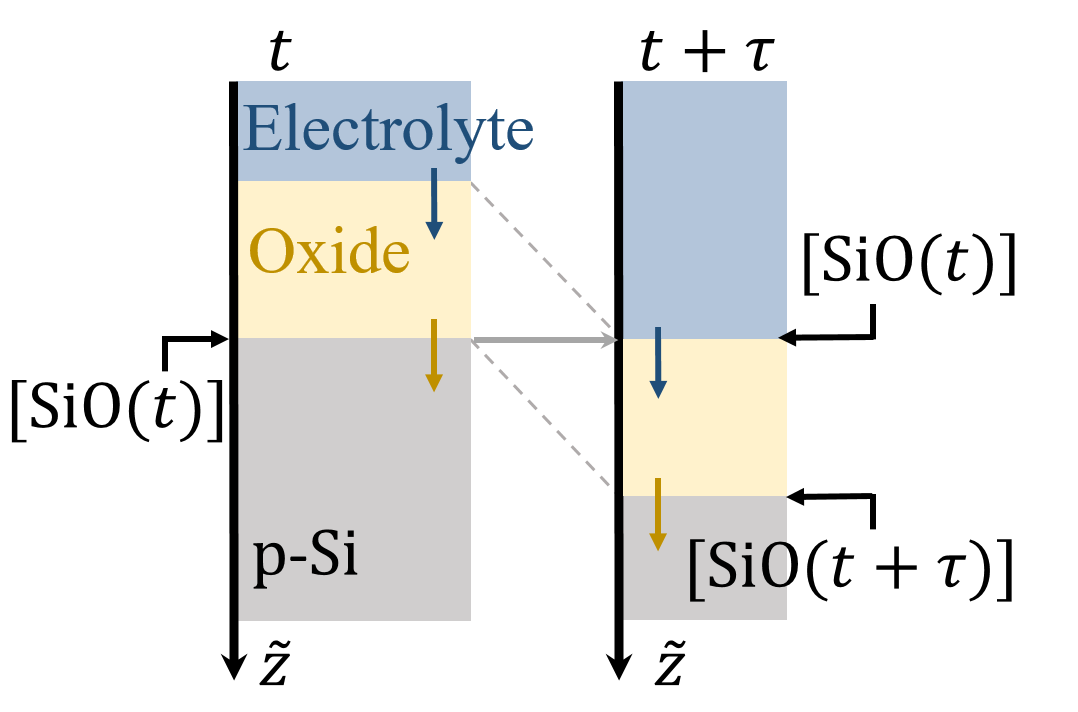}
\caption{
Sketch of the Si/oxide/electrolyte interface at $t = 0$ and $t = \tau$, as assumed in the simpmlified model.
}
\label{fig:tau_explain}
\end{figure}

The etching rate is given by eq.~(\ref{eq:etching}).
In the simplified model, we express the number density of the partially oxidized defects at the oxide/electrolyte interface through its number density at the bottom layer.  
As seen in section~\ref{sec:Results of the Full Model},  SiO species hardly diffuse in the oxide. 
Rather, in the stationary (laboratory) frame, they practically remain at the $z$ position where they were formed. 
Thus, they reach the top layer when the oxide layer above them is etched, i.e., $\tau$ time units later. 
When $[\rm SiO({\it t})]$ denotes the number density of SiO at the bottom layer at time $t$, the defect number density at the top layer is, thus, well approximated by $[{\rm SiO}(t-\tau)]$, and, consequently, the number density of $\rm SiO_2$ at the top layer at time $t$ is equal to $1-[{\rm SiO}(t-\tau)]$.
Furthermore, the results of the full model show that $\tau$ does not change much during an oscillation. 
In the simplified model, we approximate $\tau$ as a constant.
The assumed changes of the oxide position and composition during a time interval of $\tau$  are sketched in Fig.~\ref{fig:tau_explain}.
Accordingly, the etching rate is expressed as
\begin{align}
v_{\rm etch}(t) &= \gamma_2( 1-[{\rm SiO}(t-\tau)]+k_{\rm SiO}[{\rm SiO}(t-\tau)]) \label{eq:v_etch},
\end{align}
where $\gamma_2$ and $k_{\rm SiO}$ are constants.

\subsection{Defect Number Density [SiO]}
The number density of partially oxidized defects at the bottom layer, $\rm [SiO]$, results from the oxidation reactions (\ref{eq:reaction1}) and (\ref{eq:reaction2}). 
We consider a differential equation that approximately describes the time evolution of [SiO] at the Si/oxide interface. 
In the full model, the effects of reactions (\ref{eq:reaction1}) and (\ref{eq:reaction2}) were considered in the moving frame.
In this moving coordinate system, the interface moves at a velocity $v_{\rm ox} = r_1/\rho_{\rm atom}$.
When we consider an infinitesimally thin volume at this moving interface, the influx and outflux of SiO due to reaction (\ref{eq:reaction1}) are equal, meaning that (\ref{eq:reaction1}) does not change [SiO].
On the other hand, the influence of (\ref{eq:reaction2}) must satisfy the second oscillation condition, namely that the number of defects decreases with increasing electric field.
In simulations of the full model, the value of [SiO]$|_{ z_{\rm b}}$ stabilizes near a constant value. 
Therefore, in this simplified model, we assume that [SiO] relaxes to a given value (the first term in eq.~(\ref{eq:dSiO}), and we define a sigmoid function (the second term) that decreases the defect fraction with increasing $E$, as does (\ref{eq:reaction2}) in the full model.

\begin{align}
\frac{{\rm d} [{\rm SiO}]}{{\rm d}t} &= \gamma_4(\gamma_5 -[{\rm SiO}(t)])- \frac{\gamma_6 [{\rm SiO}(t)]}{1+\exp(\gamma_7 (\gamma_8 -E))} \label{eq:dSiO}.
\end{align}

\section{Results of the Simplified Model} \label{section:ResSimplModel}
In this section, we present the simulations obtained using the simplified model and examine the oscillation mechanism in more detail through linear stability analysis. 

\subsection{Potential Dependence}

\begin{figure}[htbp] 
\centering
\includegraphics[width=0.45\textwidth]{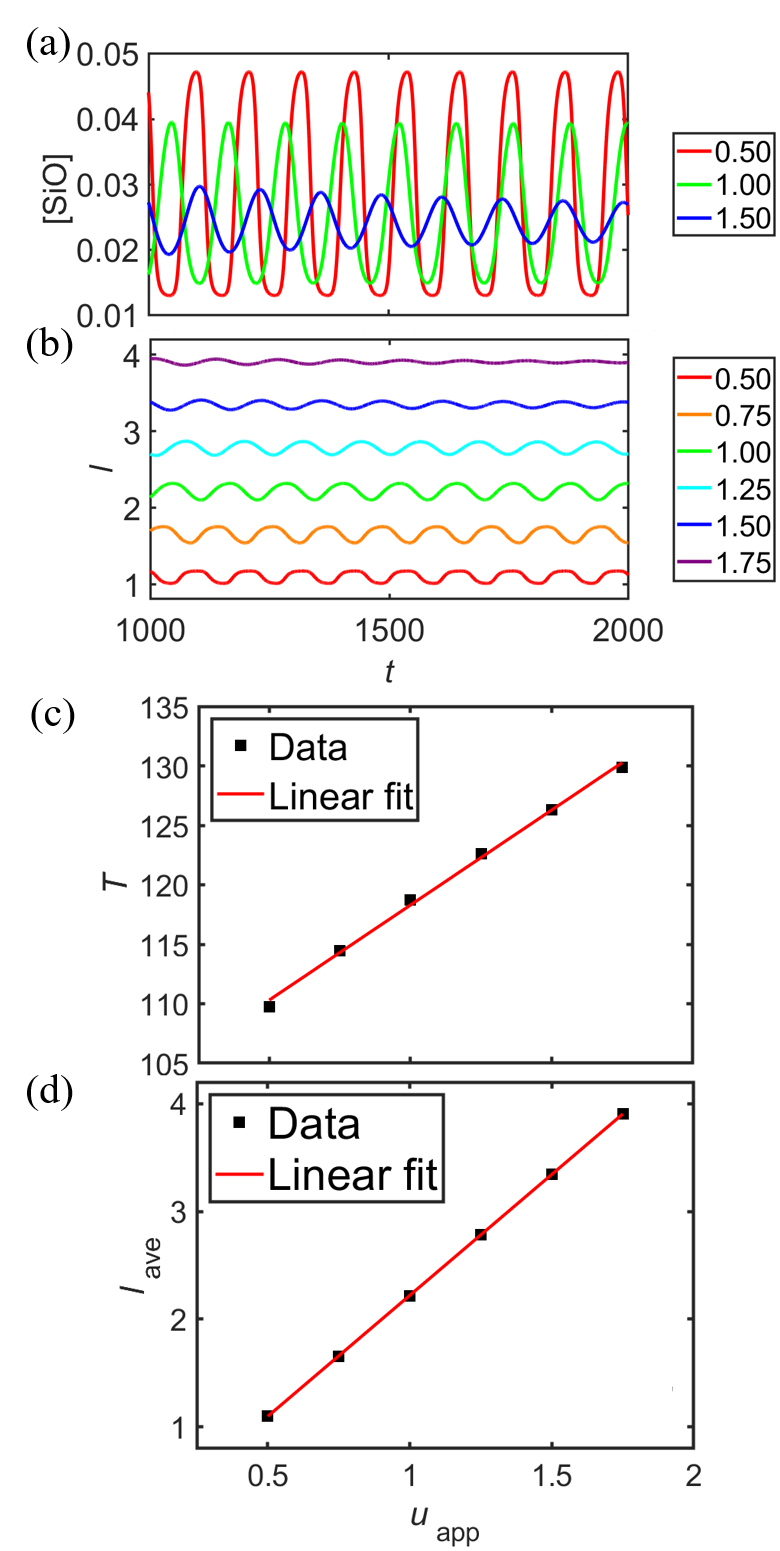}
\caption{
Simulation results of the simplified model eqs.~(\ref{eq:v_diff}) and~(\ref{eq:dSiO}). (a) time series of the SiO number density, (b) time series of the oxide-layer thickness, (c)  oscillation period, and (d) time-averaged oxide-layer thickness as a function of the applied voltage.
The different colors in (a) and (b) represent different values of the applied voltage between 0.5 and 1.75.
In (a), only three curves are shown to avoid the overlap of the data.
}
\label{fig:multi_simple}
\end{figure}

Fig.~\ref{fig:multi_simple} shows simulation results obtained with the simplified model eq.~(\ref{eq:v_etch}) and eq.~(\ref{eq:dSiO}) under fixed potential conditions between $u_{\rm app}=0.5$ and 1.75 with $\tau= 50$. 
The time series of the number density of the defects at the oxide/electrolyte interface and the oxide-layer thickness are plotted in (a) and (b), respectively.
Obviously, this simplified model also captures the oscillations.
The dependence of the oscillation period and the average oxide thickness on the applied voltage is plotted in (c) and (d), respectively.
As in the full model, both the period of the oscillations and the oxide depth increase linearly with the potential. 
However, in the simplified model, the period increases by less than 20 $\%$ while the thickness becomes nearly four times as thick. 
In contrast, the full model predicts an increase in oscillation period by a factor of 2 in a voltage interval in which the average thickness does not fully double (cf. Fig.~\ref{fig:multi_current_thickness}). 
This discrepancy arises from the fact that in the simulations, we used the same value of $\tau$, whereas in the real system, the greater thickness at higher values of $u_{\rm app}$ also implies a larger value of the delay, $\tau$. 
In positive terms, the period is again about twice as long as the delay time, $\tau$, namely $(2.4 \pm 0.2) \times \tau$, which is an inherent feature of the mechanism and, as already mentioned above, in good agreement with experimental data \cite{Duportal_2024}.

\begin{figure}
\centering
\includegraphics[width=0.48\textwidth]{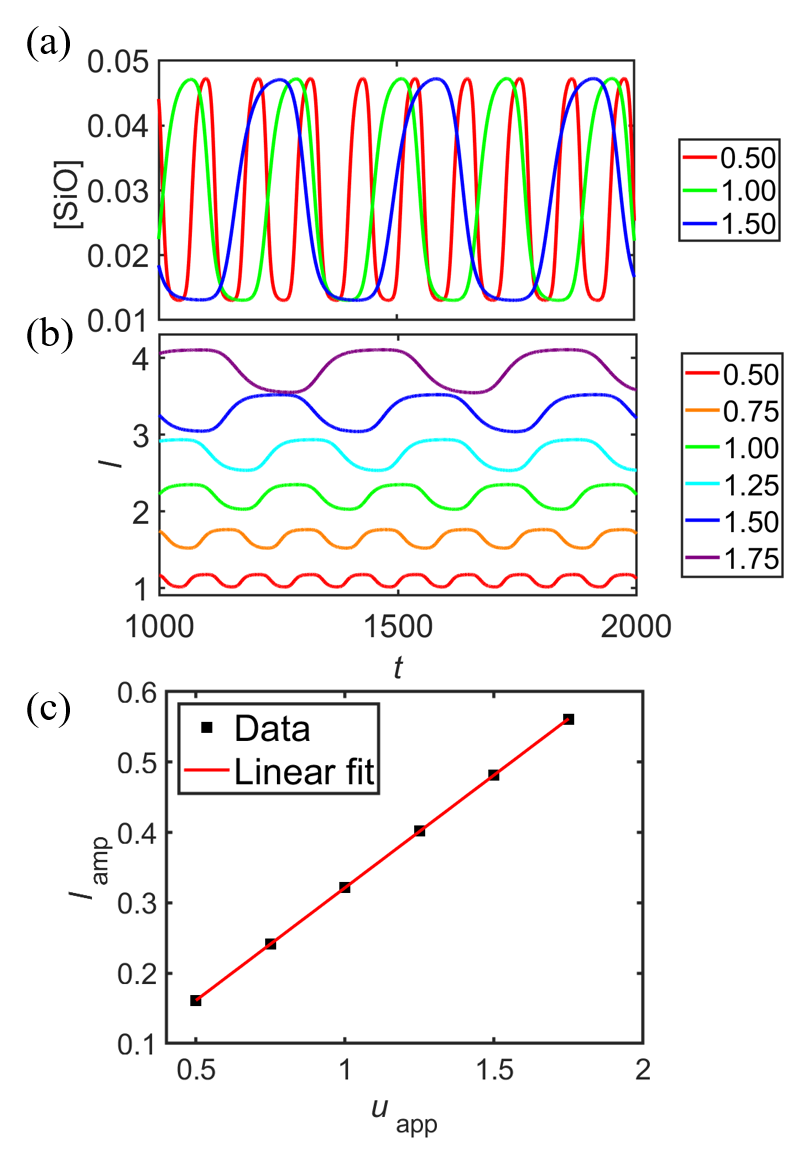}
\caption{
Simulation results of the simplified model eqs.~(\ref{eq:v_diff}) and~(\ref{eq:dSiO}) with $\tau$ being adjusted for each value of $u\rm_{app}$ to the thickness of the oxide layer as obtained with a fixed value of $\tau$ (see Fig.~\ref{fig:multi_simple}). (a) the time series of SiO number density, (b) time series of the oxide-layer thickness, and (c)  time-averaged oxide-layer thickness as a function of the applied voltage.
The different colors in (a) and (b) represent different applied potentials between 0.5 and 1.75. 
In ascending order of the potential, $\tau$ is set to 50, 75, 100, 126, 151, and 177.
In (a), only three curves are shown to avoid the overlap of the data.
}
\label{fig:multi_simple_flex}
\end{figure}

To support this interpretation, in Fig.~\ref{fig:multi_simple_flex}, we show simulation results in the same potential range as in Fig.~\ref{fig:multi_simple} but with adjusted values of $\tau$, which were taken to be proportional to the average oxide depth as obtained in simulations with a constant value of $\tau$ (see~Fig.~\ref{fig:multi_simple}~(d)).
The selected values of $\tau$ in ascending order of potential are 50, 75, 100, 126, 151, and 177.
Again, the defect number density and the oxide thickness time series are plotted in plates (a) and (b), respectively.
Here, the oscillation amplitude of the defect number density stays at a similar value in the potential range, and the oscillation amplitude of the oxide thickness increases with increasing potential.
This relation is plotted in (c). 

\subsection{Linear Stability Analysis}
Having seen that the simplified model qualitatively captures the Si oscillations, in this section, we substantiate the interpretation of the oscillation mechanism, in particular the essential role of the delay time, $\tau$, by means of a linear stability analysis.

For simplicity, we rename the two variables as follows.
\begin{align}
\begin{pmatrix}
 l(t) \\ [{\rm SiO}(t)] 
\end{pmatrix}
\equiv
\begin{pmatrix}
 x(t) \\ y(t) 
\end{pmatrix}
\equiv 
{\bm x}(t)
\end{align}
The governing equations are then given by
\begin{align}
\dot{x(t)}  &= \frac{\gamma_1 u_{\rm app}}{x(t)} - \gamma_2(1+\gamma_3 y(t-\tau))  \label{eq:dxdt}\\
\dot{y(t)}  &= \gamma_4(\gamma_5 - y(t)) - \frac{\gamma_6 y(t)}{1+\exp \left( \gamma_7 \left( \gamma_8 -\frac{u_{\rm app}}{x(t)} \right) \right)}. \label{eq:dydt}
\end{align}
At a fixed point, $y(t)$ and $y(t-\tau)$ are equal. 
Therefore, it is straightforward to numerically solve for the fixed point, $(x^*, y^*)$, determined by
$\dot{x(t)}=0$ and $\dot{y(t)}=0$.
We linearize the system around the fixed point and analyze the linear stability for small perturbations.
With the following definitions of $f(x,y)$ and $g(x,y)$ 
\begin{align}
f(x,y)  &\equiv  \frac{\gamma_1 u_{\rm app}}{x} - \gamma_2(1+\gamma_3 y) \\
g(x,y)  &\equiv  \gamma_4(\gamma_5 - y) - \frac{\gamma_6 y}{1+\exp \left( \gamma_7 \left( \gamma_8 -\frac{u_{\rm app}}{x} \right) \right)},
\end{align}
the partial derivatives are given by
\begin{align}
\frac{\partial f}{\partial x}  &=  - \frac{\gamma_1 U_{\rm eff}}{x^2} \\
\frac{\partial f}{\partial y}  &=   - \gamma_2 \gamma_3 \\
\frac{\partial g}{\partial x}  &=  \frac{\gamma_6 y \frac{\gamma_7 u_{\rm app}}{x^2} \exp(\gamma_7(\gamma_8- \frac{U_{\rm eff}}{x}))}{\left(1+\exp \left( \gamma_7 \left( \gamma_8 -\frac{u_{\rm app}}{x} \right) \right) \right)^2} \\
\frac{\partial g}{\partial y}  &=  -\gamma_4 -\frac{\gamma_6}{1+\exp \left( \gamma_7 \left( \gamma_8 -\frac{u_{\rm app}}{x} \right) \right)}.
\end{align}
The linearized system can then be expressed as 
\begin{align}
\includecomment{}{{\bm x}}(t) = A{\bm x}(t) + B {\bm x}(t-\tau),
\end{align}
where $A$ and $B$ are defined by
\begin{align}
A
\equiv
\begin{pmatrix}
\frac{\partial f}{\partial x} & 0 \\
\frac{\partial g}{\partial x} & \frac{\partial g}{\partial y}
\end{pmatrix}_{(x^*, y^*)},
\qquad
B
\equiv
\begin{pmatrix}
0 & \frac{\partial f}{\partial y} \\
0  & 0 
\end{pmatrix}_{(x^*, y^*)}.
\end{align}

If the oscillation mechanism discussed above is correct, the stability of the fixed point should depend on the delay time, $\tau$, losing its stability for increasing $\tau$.
This can be verified by means of a Laplace transform.
With the following definition of the functions $p(s,\pm1)$ 
\begin{align}
p(s,\pm1) 
\equiv
\det (sI - (A \pm B)), \label{eq:linearstab}
\end{align}
the fixed point is stable if 1. the real parts of the roots of $p(s_+,1)=0$ are both negative and 2. at least one of the real parts of the roots of $p(s_-,-1)=0$ is positive \cite{Dey_2017stability}.
The numerical solution of eq.~(\ref{eq:linearstab}) shows that this system satisfies the delay-dependent stability conditions.
\begin{align}
{\rm Re}(s_+^1)< 0 \land {\rm Re}(s_+^2)< 0  \\
{\rm Re}(s_-^1) > 0  \lor {\rm Re}(s_-^2) > 0
\end{align}
For a certain value of $\tau$, ${\rm Re}(s_{1,2}) = 0$, and the roots $s_i$ are purely imaginary. This means that a Hopf bifurcation occurs and a limit cycle is born.
Correspondingly, the amplitude exhibits a square root dependence as a function of the voltage, as seen in Fig.~\ref{fig:multi_current_thickness} (d).
The existence of a Hopf bifurcation in this system is also confirmed in the experiments \cite{Schoenleber2016}.

\section{Conclusion} \label{section:Conclusion}
Starting from ideas of the point defect model that describes the motion of point defects in passive films on metal surfaces, we derived a mathematical model for the formation and dissolution of oxide layers during the anodic polarization of silicon in fluoride-containing electrolytes. Our minimal model incorporates just two defects, substoichiometric SiO and oxygen valencies, SiO$^{2+}$, whose temporal evolution is described by mass balance equations coupled to Poisson's equation in the oxide layer. Oscillations arise when three conditions are met: Firstly, the more substoichiometric SiO the oxide layer contains, the faster it is etched. Secondly, the fraction of substoichiometric oxide formed at the Si/oxide interface is the higher, the lower the electric field at the interface is. In other words, a large electric field yields more fully oxidized SiO$_2$. Thirdly, the time delay between the formation of oxide at the Si/oxide interface and its etching at the oxide/electrolyte interface must be sufficiently long. This third condition creates positive feedback, which, in conjunction with the negative feedback loops of the first two conditions, enables oscillations. 
The numerical results of the model agree well with experimentally observed data. In particular, the dependence of current, oxide-layer thickness, and period on the applied voltage is captured under potentiostatic and potentiodynamic conditions. Furthermore, the oxide layer is etched approximately twice during one oscillatory cycle in simulations and experiments. 

The oscillation mechanism was validated using a simplified two-variable time-delay differential model that contained only the essential features of the oscillation mechanisms. The simplified model allowed us to perform a linear stability analysis that  further confirmed the interpretation of the oscillation mechanism. 

We believe that the model provides a solid starting point for further investigations in various directions. First, oscillations were observed during anodic polarization of many semiconductor and metal electrodes, most of which are not yet understood. The oscillation mechanism discovered here is not related to specific properties of silicon electrochemistry. Rather, it seems to be likely that the three components of the mechanism are also present in other electrode materials and could therefore explain the occurrence of oscillations in various systems. Second, the model appears to provide the basic properties of the electrochemistry of oxide-covered silicon electrodes. Extensions of the model seem to be straightforward and should uncover further fundamental properties of Si electrochemistry. Among them is a better description of the composition of oxide by allowing electrolyte species to penetrate into the oxide, the formation of pores when the electric field strength becomes large, as considered in the current-burst model, or charge-transfer reactions within the oxide, that seem to dominate the dynamics in the negative differential resistance region (II) in cyclic voltammograms (cf.  Fig.~\ref{fig:CV_diff_speed}). Thirdly, extending the model to include spatial directions parallel to the electrode surface should provide information about properties of the spatial coupling of different locations of the electrode through valence-band holes, and their role in spatial pattern formation. It has been speculated that this coupling is responsible for the formation of unusual patterns such as chimera states or frequency clusters. Therefore, expanding this research will establish an important link between the numerous theoretical studies on these patterns and their experimental realization. 

\section{Acknowledgements}
We thank M. Duportal for fruitful discussions. This project was funded by the Deutsche Forschungsgemeinschaft (DFG, German
Research Foundation, project KR1189/18-2).

\section*{Declaration of interests}
The authors declare no competing interests.

\section{Appendix} \label{Appendix}
\subsection{Experimental Conditions} \label{sec:Experimental Conditions}
The electrochemical experiment data shown in this paper were measured with a three-electrode cell.
The working electrode is made of p-doped 1-10 $\rm \Omega cm^2$ silicon with a back contact by thermally evaporated aluminum on the back side and annealed at 250 ${}^\circ$C for 15 minutes.
The reference electrode is saturated $\rm Hg|HgSO_4$, and the counter electrode is made by platinum wire.
The electrolyte comprises 60 mM $\rm NH_4F$ and 142 mM $\rm H_2SO_4$ at room temperature.
The electrolyte was deaerated by bubbling with argon for 20 minutes.
During the measurement, the electrolyte was stirred by a magnetic stirrer.

\subsection{Parameters of the Full Model} \label{section:Parameters of the Full Model}
The follow parameters were used for the calculation presented in section~\ref{sec:Results of the Full Model}.
\subsubsection{Natural Parameters}
\begin{align}
q                 &= 1.602 \times 10^{-19} &[\rm C] \nonumber \\
k_{\rm B}T        &= 1.381\times 10^{-23} \times 300  &[\rm J] \nonumber \\ 
\epsilon_0        &= 8.85 \times 10^{-12}  &[\rm F/m] \nonumber \\
\epsilon_{\rm sc} &= 11.7  &[-] \nonumber \\
\epsilon_{\rm ox} &= 3.8  &[-] \nonumber \\
\rho_0            & = 1 \times 10^{21}  &[\rm 1/m^3] \nonumber \\
\rho_{\rm atom}   & = 2.66 \times 10^{28}  &[\rm 1/m^3] \nonumber \\
l_{\rm m}         & = 3.135 \times 10^{-10}  &[\rm m] \nonumber 
\end{align}
\subsubsection{Calculated Parameters}
\begin{align}
L_{\rm D}              &= \frac{\sqrt{k_{\rm B}T  \epsilon_0 \epsilon_{\rm sc} / \rho_0}}{q}  &[\rm m] \nonumber \\
k_{v_{\rm ox}}   &= \frac{1}{ \rho_{\rm atom}} &[\rm m^3] \nonumber \\
D_{\rm SiO^{2+}} &= \frac{\mu_{\rm SiO^{2+}} k_{\rm B}T}{2q} &[\rm m^2/s] \nonumber \\
\end{align}
\subsubsection{Assumed Parameters}
\begin{align}
L_0                &= 4  \times 10^{-9}  &[\rm m] \nonumber \\
\mu_{\rm SiO^{2+}} &= 4  \times 10^{-17}  &[\rm m^2/(V s)] \nonumber \\
D_{\rm SiO} &= 1\times 10^{-20}  &[\rm m^2/ s] \nonumber \\
C_{\rm sc}         &= 2.85 \times 10^{-2} &[\rm F/m^2] \nonumber \\
k_{v_{\rm etch}}   &= 4.8 \times 10^{-11} / \rho_{\rm atom} &[\rm m^4/s] \nonumber \\
k_{\rm SiO}        &= 6 &[-] \nonumber \\
k_1                &= 6 \times 10^{-6}/(2q \rho_{\rm atom}) &[\rm  m^2/(Vs)] \nonumber \\
\alpha_1           &= 7 \times 10^{3} &[\rm m/V] \nonumber \\
k_2                &= 1 \times 10^{-4}/(2q \rho_{\rm atom}) &[\rm 1/(ms)] \nonumber \\
\alpha_2           &= 66 &[1/\rm V] \nonumber \\
\alpha_3           &= 5/11 &[\rm V] \nonumber \\
k_{\rm leak}       &=  1 \times 10^{-5} \times \rho_{\rm atom} &[\rm 1/m^3] \nonumber 
\end{align}
\subsubsection{Adjustable Parameters} 
\begin{align}
n &= 25 &[-] \nonumber  \\
A &= 1   &[\rm m^2] \nonumber  \\
{\rm d}t &= 1 \times 10^{-5}  &[\rm s] \nonumber 
\end{align}
\subsection{Parameters of the Simplified Model} \label{AppendixParaSimplifiedM}
\begin{align}
\gamma_1  &= 0.5  &[-] \nonumber  \\
\gamma_2  &= 0.2  &[-] \nonumber \\
\gamma_3  &= 5    &[-] \nonumber \\
\gamma_4  &= 0.2  &[-] \nonumber \\
\gamma_5  &= 0.6  &[-] \nonumber \\
\gamma_6  &= 10   &[-] \nonumber \\
\gamma_7  &= 50   &[-] \nonumber \\
\gamma_8  &= 0.45 &[-] \nonumber 
\end{align}

\bibliographystyle{ieeetr}

\begin{thebibliography}{10}

\bibitem{Zhang2001}
X.~G. Zhang, {\em {Electrochemistry of silicon and its oxide}}.
\newblock New York: Kluwer Academic/Plenum Publishers, 2001.

\bibitem{Turner1958}
D.~R. Turner, ``{Electropolishing Silicon in Hydrofluoric Acid Solutions},''
  {\em J. Electrochem. Soc.}, vol.~105, no.~7, pp.~402--408, 1958.

\bibitem{Wiehl2021}
J.~C. Wiehl, M.~Patzauer, and K.~Krischer, ``Birhythmicity, intrinsic
  entrainment, and minimal chimeras in an electrochemical experiment,'' {\em
  Chaos: An Interdisciplinary Journal of Nonlinear Science}, vol.~31,
  p.~091102, 09 2021.

\bibitem{Tosolini2019}
A.~Tosolini, M.~Patzauer, and K.~Krischer, ``{Bichaoticity induced by inherent
  birhythmicity during the oscillatory electrodissolution of silicon},'' {\em
  Chaos: An Interdisciplinary Journal of Nonlinear Science}, vol.~29,
  p.~043127, 4 2019.

\bibitem{Miethe2009}
I.~Miethe, V.~Garc{\'{i}}a-Morales, and K.~Krischer, ``{Irregular Subharmonic
  Cluster Patterns in an Autonomous Photoelectrochemical Oscillator},'' {\em
  Physical Review Letters}, vol.~102, pp.~94 -- 101, 5 2009.

\bibitem{Schoenleber2014}
K.~Sch{\"{o}}nleber, C.~Zensen, A.~Heinrich, and K.~Krischer, ``{Pattern
  formation during the oscillatory photoelectrodissolution of n-type silicon:
  Turbulence, clusters and chimeras},'' {\em New Journal of Physics}, vol.~16,
  p.~63024, 6 2014.

\bibitem{Schmidt2014}
L.~Schmidt, K.~Sch{\"{o}}nleber, K.~Krischer, and V.~Garc{\'{i}}a-Morales,
  ``{Coexistence of synchrony and incoherence in oscillatory media under
  nonlinear global coupling},'' {\em Chaos (Woodbury, N.Y.)}, vol.~24,
  p.~013102, 3 2014.

\bibitem{Patzauer2021}
M.~Patzauer and K.~Krischer, ``Self-organized multi-frequency clusters in an
  oscillating electrochemical system with strong nonlinear coupling,'' {\em
  Physical Review Letters}, vol.~126, 5 2021.

\bibitem{Cattarin2000}
S.~Cattarin, I.~Frateur, M.~Musiani, and B.~Tribollet, ``{Electrodissolution of
  p-Si in Acidic Fluoride Media Modeling of the Steady State},'' {\em Journal
  of The Electrochemical Society}, vol.~147, no.~9, p.~3277, 2000.

\bibitem{Chazalviel1998}
J.-N. Chazalviel, C.~da~Fonseca, and F.~Ozanam, ``{In Situ Infrared Study of
  the Oscillating Anodic Dissolution of Silicon in Fluoride Electrolytes},''
  {\em Journal of The Electrochemical Society}, vol.~145, no.~3, p.~964, 1998.

\bibitem{Cattarin1998a}
S.~Cattarin, J.-N. Chazalviel, C.~da~Fonseca, F.~Ozanam, L.~M. Peter,
  G.~Schlichth{\"{o}}rl, and J.~Stumper, ``{In Situ Characterization of the
  p-Si/NH$_4$F Interface during Dissolution in the Current Oscillations
  Regime},'' {\em Journal of the Electrochemical Society}, vol.~145, no.~2,
  pp.~498--502, 1998.

\bibitem{Aggour1995}
M.~Aggour, M.~Giersig, and H.~J. Lewerenz, ``{Interface condition of n-Si (111)
  during photocurrent oscillations in NH 4 F solutions},'' {\em Journal of
  Electroanalytical Chemistry}, vol.~383, no.~1-2, pp.~67--74, 1995.

\bibitem{Lehmann1996}
V.~Lehmann, ``{On the Origin of Electrochemical Oscillations at Silicon
  Electrodes},'' {\em Journal of The Electrochemical Society}, vol.~143, no.~4,
  p.~1313, 1996.

\bibitem{Ozanam1992}
F.~Ozanam, J.-N. Chazalviel, A.~Radi, and M.~Etman, ``{Resonant and nonresonant
  behavior of the anodic dissolution of silicon in fluoride media: {\{}An
  impedance Study{\}}},'' {\em J. Electrochem. Soc.}, vol.~139/9, p.~2491,
  1992.

\bibitem{Grzanna2000a}
J.~Grzanna, H.~Jungblut, and H.~J. Lewerenz, ``{A model for electrochemical
  oscillations at the Si vertical bar electrolyte contact Part I. Theoretical
  development},'' {\em J Electroanal Chem}, vol.~486, no.~2, pp.~181--189,
  2000.

\bibitem{Schoenleber2012}
K.~Sch{\"{o}}nleber and K.~Krischer, ``{High-amplitude versus low-amplitude
  current oscillations during the anodic oxidation of p-type silicon in
  fluoride containing electrolytes},'' {\em Chem. Phys. Chem.}, vol.~13,
  pp.~2989--2996, 8 2012.

\bibitem{Smith1992}
R.~L. Smith and S.~D. Collins, ``Porous silicon formation mechanisms,'' {\em
  Journal of Applied Physics}, vol.~71, pp.~R1--R22, 04 1992.

\bibitem{Macdonald_1992}
D.~D. Macdonald, ``The point defect model for the passive state,'' {\em Journal
  of The Electrochemical Society}, vol.~139, p.~3434, dec 1992.

\bibitem{Lewerenz1988}
H.~J. Lewerenz, J.~Stumper, and L.~M. Peter, ``{Deconvolution of charge
  injection steps in quantum yield multiplication on silicon},'' {\em Physical
  Review Letters}, vol.~61, pp.~1989--1992, 10 1988.

\bibitem{Sze_Ng_2006}
S.~Sze and K.~Ng, {\em Physics of Semiconductor Devices. 3rd Edition}.
\newblock Hoboken, New Jersey: John Wiley and Sons, Inc., 2006.

\bibitem{Duportal_2024}
M.~Duportal, A.~Tosolini, J.~C. Wiehl, Y.~Murakami, and K.~Krischer,
  ``Application of the point defect model to the oscillatory anodic oxidation
  of illuminated n-type silicon in the presence of fluoride ions using
  electrochemical impedance spectroscopy,'' {\em Journal of The Electrochemical
  Society}, vol.~171, p.~086505, aug 2024.

\bibitem{Salman2019}
M.~M. Salman, M.~Patzauer, D.~Koster, F.~{La Mantia}, and K.~Krischer,
  ``{Electro-oxidation of p-silicon in fluoride-containing electrolyte: a
  physical model for the regime of negative differential resistance},'' {\em
  The European Physical Journal Special Topics}, vol.~227, pp.~2641--2658, 4
  2019.

\bibitem{Dey_2017stability}
D.~Rajeeb, R.~Goshaidas, and E.~B. Valentina, {\em Stability and Stabilization
  of Linear and Fuzzy Time-Delay Systems}.
\newblock Singapore: Springer, 2017.

\bibitem{Schoenleber2016}
K.~Sch{\"{o}}nleber, M.~Patzauer, and K.~Krischer, ``{A comparison of modeling
  frameworks for the oscillatory silicon electrodissolution},'' {\em
  Electrochimica Acta}, vol.~210, pp.~346--351, 2016.

\end{thebibliography}

\end{document}